\documentclass[%
 prd,
 twocolumn,
 superscriptaddress,
 numerical,
 showpacs,
 amsmath,amssymb,
 aps,
nofootinbib,
preprintnumbers,
longbibliography,
 floatfix
]{revtex4-1}
\usepackage[utf8]{inputenc}
\usepackage{amsmath}
\usepackage{amsfonts}
\usepackage{amssymb}
\usepackage{color}
\usepackage{graphicx}
\usepackage{epstopdf}
\usepackage{verbatim}
\usepackage{cancel}
\usepackage{siunitx}
\usepackage{braket}
\usepackage[colorlinks,linkcolor=blue,citecolor=blue]{hyperref}

\definecolor{DeepPink}{rgb}{0.796,0.004,0.384}
\definecolor{RoyalBlue}{rgb}{0.02,0.016,0.667}
\definecolor{Teja}{rgb}{0.65,0.20,0.15}
\definecolor{Berry}{rgb}{0.60,0.06,0.34}

\hypersetup{
    colorlinks=true,
    linkcolor=blue,
    urlcolor=cyan,
    citecolor=cyan,}

\usepackage{fancyvrb}
\usepackage{subfig}
\numberwithin{equation}{section}

\usepackage{multirow}
\usepackage{tikz}
 
\usepackage{array}

\begin{document}

\title{Chiral Magnetic Effect enhancement at lower collision energies}

\author{Sebastian Grieninger}
\email{sebastian.grieninger@stonybrook.edu}
\affiliation{Center for Nuclear Theory, Department of Physics and Astronomy,
Stony Brook University, Stony Brook, New York 11794–3800, USA}
\author{Sergio Morales-Tejera}
\email{sergio.morales@e-uvt.ro}
\affiliation{Department of Physics, West University of Timisoara, 
Bd. Vasile Parvan 4, Timisoara 300223, Romania}
\author{Pau G. Romeu}
\email{pau.garcia@uam.es}
\affiliation{Instituto de F\'isica Te\'orica UAM/CSIC, c/Nicol\'as Cabrera 13-15, Universidad Aut\'onoma de Madrid, Cantoblanco, 28049 Madrid, Spain} \affiliation{  Departamento de F\'isica Te\'orica, Universidad Aut{\'o}noma de Madrid, Campus de Cantoblanco, 28049 Madrid, Spain}

\preprint{IFT-UAM/CSIC-25-24}
\date{\today}

\begin{abstract}

We extend previous holographic studies of the Chiral Magnetic Effect (CME) by incorporating a time-dependent magnetic field. Various magnetic field profiles proposed in the literature are implemented, and their impact on the CME signal is analyzed in both static and expanding backgrounds. Interestingly, the integrated chiral magnetic current can exhibit a non-monotonic dependence on the collision energy. Our results suggest that the CME signal is enhanced at collision energies below $\sqrt{s}=200$ GeV.  In addition, we derive a quasi-equilibrium formula for the chiral magnetic effect in the expanding background that is valid at late times.

\end{abstract}

\maketitle

\newpage
{ 
\hypersetup{linktocpage=true}
}

\newpage

\section{Introduction}
The quark-gluon plasma (QGP) formed in heavy-ion collisions provides an excellent window into the properties of QCD matter under extreme conditions. The fast moving ions in non-central collisions generate sizable magnetic fields~\cite{Skokov:2009qp,Bzdak:2011yy,Voronyuk:2011jd,Deng:2012pc,Tuchin:2013apa,McLerran:2013hla,Tuchin:2013ie,Tuchin:2014hza,Gursoy:2014aka,Li:2016tel,Pasechnik:2016wkt,Mayer:2024kkv}, while the off-central collisions imprint a large angular momentum onto the plasma, causing it to rotate. The magnetic field strength can reach magnitudes of $eB \approx 10m_\pi^2$, where
$m_\pi \approx 0.14$ GeV is the pion mass, in ultrarelativistic heavy ion collisions at the Relativistic Heavy Ion Collider (RHIC) at energies of $\sqrt{s_{NN}}$ = 200 GeV. These values are even exceeded at the Large Hadron Collider (LHC). Similarly, the off-central collisions  The vorticity of the plasma can be accessed through the decay of the hyperons, and is estimated to be of the order of $10$ MeV \cite{STAR:2017ckg}. Combined with the chiral imbalance of the plasma, both the magnetic field and angular velocity can give rise to a variety of anomaly induced transport phenomena, as was recently reviewed in Refs.~\cite{Kharzeev:2015znc,Adhikari:2024bfa}.

Among these, the Chiral Magnetic Effect (CME) stands out as one of the most fascinating manifestations~\cite{Kharzeev:2004ey,Kharzeev:2007jp,Fukushima:2008xe}, and has received most of the theoretical and experimental attention. The CME is the phenomenon where an electric current is generated along the direction of the magnetic field due to the imbalance between left- and right-handed particles (for a recent review see \cite{Kharzeev:2024zzm} and references therein). The chiral imbalance has been argued to arise from topological fluctuations of the non-abelian gauge fields in the QGP \cite{Kharzeev:2004ey}. We refer to the electric current resulting from the CME as the chiral magnetic current. Similarly, the magnetic field combined with a particle/anti-particle imbalance gives rise to the Chiral Separation Effect~\cite{Son:2004tq,Son:2007ny}, while a helicity imbalance leads to the Helical Separation Effect~\cite{Ambrus:2023erf}. The coherent interplay between the CME and CSE further gives rise to the Chiral Magnetic Wave~\cite{Kharzeev:2010gd,Burnier:2011bf}. Note also the recently discovered transport phenomena of Ref.~\cite{Ammon:2020rvg}. We refer the reader to Refs.~\cite{Erdmenger:2008rm,Son:2009tf,Jiang:2015cva,Morales-Tejera:2024mtx,Morales-Tejera:2024uzg} for the rotational counterparts of the anomaly induced transport phenomena. Taken together, these provide a unique opportunity to observe macroscopic effects of quantum anomalies in quantum chromodynamics (QCD), and to probe indirectly the non-trivial topology of the non-abelian gauge fields. In this work, we focus on the chiral magnetic effect phenomenology in the context of heavy ion collisions.

The existence of the CME has been experimentally verified in condensed matter experiments \cite{Li:2014bha,li2015giant,xiong2015evidence,PhysRevX.5.031023}. The measurement of this effect in heavy ion collisions appears to be more elusive, with preliminary searches at both RHIC and LHC, and a subsequent dedicated experiment at RHIC known as the isobar runs. The blind analysis of the isobar runs by the Star Collaboration \cite{STAR:2021mii} concluded that there was no CME signal according to the pre-defined criteria of the experiment. These criteria were later re-examined and a post-blind analysis was presented in Refs. \cite{Kharzeev:2022hqz,Lacey:2022plw}, with results consistent with a finite CME signal. However, this conclusion is not definitive, and further experimental analyses are being carried out. In this paper we consider the theoretical uncertainties arising from the time dependence of the magnetic field, which may help to understand the experimental results.

The magnitude of the chiral magnetic current strongly depends on the strength of the magnetic field. However, the exact time dependence of the magnetic field during heavy-ion collisions is still debated in the literature. As was recently suggested in~\cite{Huang:2022qdn}, the presence of a conducting medium (such as the QGP itself) can significantly extend the lifetime of these magnetic fields through a mechanism analogous to Faraday induction. As the external magnetic field begins to decay, it induces electric currents within the conducting QGP, which in turn generate secondary magnetic fields that partially counteract the decay of the primary field. The authors~\cite{Huang:2022qdn} argue that this interplay between the decaying external field and the response of the medium leads to a much more complex time evolution than previously thought. 

The specific temporal profile of the magnetic field is not just a technical detail but a fundamental factor that directly influences the magnitude and observability of the chiral magnetic current. For this reason, we perform a comprehensive analysis of the chiral magnetic current with a variety of magnetic field profiles. The inherent non-equilibrium nature of the CME poses a challenge to its simulation within lattice QCD, although recent studies in this direction are promising \cite{Buividovich:2024bmu,Brandt:2024fpc,Brandt:2025now}. Alternatively, the AdS/CFT correspondence naturally provides a framework for addressing the real-time evolution of the chiral magnetic current in a strongly coupled  out-of-equilibrium plasma. Anomalous transport phenomena, and in particular the CME, have been extensively studied in the holographic literature, see Refs. \cite{Gynther:2010ed,Lin:2013sga,Jimenez-Alba:2014iia,Jimenez-Alba:2015awa,Ammon:2016fru,Landsteiner:2017lwm,Fernandez-Pendas:2019rkh,PhysRevD.100.126024,Ghosh:2021naw,Cartwright:2021maz,Rai:2024qmx,Grieninger:2023myf,Grieninger:2023wuq} and references therein, and we will also follow this holographic approach. However, the previous holographic studies of the CME are restricted to a constant, or Bjorken expanding\footnote{In \cite{Cartwright:2021maz,Grieninger:2023myf} the authors studied a Bjorken expanding plasma where the magnetic field decays inversely proportional to the proper time due to dilution.}, magnetic fields. In this work, we overcome this limitation and implement various time-dependent magnetic fields, subsequently studying the time evolution of the chiral magnetic current in these cases.

In particular, we systematically investigate how different temporal profiles of the magnetic field affect the CME signal, both in static and expanding backgrounds. For the time dependence of the magnetic field, we choose the profiles that are currently used as state-of-the-art in the literature, as a function of the energy of the collision, closely following the approach of Ref.~\cite{Cartwright:2021maz}. The profiles are based on detailed calculations incorporating factors such as the medium's electrical conductivity, the collision energy, and the impact parameter. By comparing these different scenarios, we can assess the sensitivity of the CME to specific details of the magnetic field evolution and identify robust features that persist across different models. Within the holographic model we make two important simplifications. First, we assume that the magnetic field is comparatively small with respect to the energy density, equivalently the temperature, of the plasma. This assumption is made more precise in Sec.~\ref{sec:model}, where its range of applicability is also discussed. We argue that this approximation is sufficient for the phenomenological studies of the QGP in heavy-ion collisions. Secondly, for the case of an expanding plasma, we consider the late-time Bjorken expanding metric \cite{Kalaydzhyan:2010iv} as the background solution, neglecting possible deviations at earlier times. We expect, from the analysis in Sec.~\ref{sec:model} that such deviations do not modify qualitatively our conclusions. Extending the late-time analysis of the equations of motion to the dynamics of the gauge fields, we are able to derive a (quasi)-equilibrium formula for the chiral magnetic current in the expanding case, 
\begin{equation}
    2\kappa^2 J^{\text{ quasi-eq}}_{\textrm{CME}} =8\alpha\mu_5(\tau) B_z(\tau)\,,
\end{equation}
in close resemblance to the known result for the equilibrium chiral magnetic current in a static plasma. The derivation of the previous formula and its validity are discussed in Sec.~\ref{sec:inilate}.

Interestingly, our results reveal a non-monotonic relationship between the time-integrated chiral magnetic current and the collision energy. This behavior results from the interplay between the increasing initial field strength and the decreasing lifetime of the magnetic field with increasing collision energy. In particular, the longer lifetime of the magnetic field at lower energies allows for more CME signal accumulation even though the initial magnetic field is weaker. Conversely, at higher energies, the extremely strong but rapidly decaying fields produce less charge accumulation. This non-monotonic behavior offers new perspectives for the interpretation of experimental data and highlights the importance of incorporating realistic time-dependent magnetic fields into theoretical models.

Our paper is organized as follows: In section~\ref{sec:model}, we introduce the holographic model used in this work and discuss the setup for the expanding and non-expanding plasmas. The different magnetic field profiles as well as the thermodynamic parameters are discussed in section~\ref{sec:III}. We then proceed to study the time dependence of the chiral magnetic current subject to time dependent magnetic fields in section~\ref{sec:results}. Finally, we summarize our findings in section~\ref{sec:conclusions} and discuss possible avenues for future research.

\section{Holographic $U(1)_V\times U(1)_A$  model}\label{sec:model}

In order to study the out-of-equilibrium response of the CME at strong coupling we consider the holographic model put forward in \cite{Gynther:2010ed}. The matter content of the model consists of two abelian gauge fields, denoted $V^\mu$ and $A^\mu$, which are dual to the vector and axial symmetries of the dual field theory. Their field strengths are denoted $F=dV$ and $F^5=dA$ respectively. The abelian contribution to the axial anomaly is captured by a Chern-Simons term. The non-abelian contribution to the anomaly which is responsible for the manifestion of the CME in QCD may be implemented in an extension of the model considered in Ref. \cite{Jimenez-Alba:2014iia}. In previous work, we have discussed its impact on the CME in Refs. \cite{Grieninger:2023myf,Grieninger:2023wuq}.

Explicitly, the action of the model is given by
\begin{align}\label{Bt:action} 
S &= \frac{1}{2\kappa^2}\int_{\mathcal{M}} d^5x\sqrt{-g}\left[R+\frac{12}{L^2}-\frac{1}{4}F^2-\frac{1}{4}F_{5}^2 \right. \nonumber \\& \left. +\frac{\alpha}{3} \epsilon^{\mu\nu\rho\sigma\tau} A_\mu\left( 3 F_{\nu\rho}F_{\sigma\tau}+F^{5}_{\nu\rho}F^{5}_{\sigma\tau}\right) \right]  +S_{GHY}
\end{align}
where $S_{GHY}$ is the Gibbons-Hawking-York boundary term to make the variational problem well defined. As usual, $\kappa$ is related to the Newton constant in five dimensions, $\mathcal{M}$ is the spacetime manifold and $L$ is the AdS scale, which we set to one without loss of generality. The Chern-Simons coupling $\alpha$ is a parameter of the model that controls the strength of the anomaly. We define the Levi-Civita tensor in terms of the Levi-Civita symbol as $\epsilon^{\mu\nu\rho\sigma\tau} = \epsilon(\mu\nu\rho\sigma\tau)/\sqrt{-g}$. Finally, we work with the mostly positive metric convention. 

Following \cite{Ghosh:2021naw}, we fix the parameters $\alpha$ and $\kappa$ by matching to the axial anomaly of QCD as well as to $3/4$ of the entropy density in the 
Stefan Boltzmann limit of three flavor QCD. In particular, this gives
\begin{equation}
    \kappa^2 = \dfrac{24 \pi^2}{19}\simeq12.5\,,\quad \alpha = \dfrac{6}{19}\simeq0.316\,.
\end{equation}
It should be noted that the holographic model describes a deconfined and chiral symmetry restored phase at all temperatures. Therefore, the results obtained at lower temperatures as we approach the phase transition of QCD become progressively less reliable. 

The equations of motion derived from \eqref{Bt:action} are given by
\begin{align}
& \nabla_\nu F^{\nu\mu}+2\alpha \epsilon^{\mu\nu\rho\sigma\tau} F_{\nu\rho}F^{(5)}_{\sigma\tau}=0, \label{eom:Maxwell}\\
& \nabla_\nu F_{(5)}^{\nu\mu}+\alpha  \epsilon^{\mu\nu\rho\sigma\tau} \left( F_{\nu\rho}F_{\sigma\tau}+F_{\nu\rho}^{(5)}F_{\sigma\tau}^{(5)} \right)=0, \label{eom:axial}\\ 
& G_{\mu\nu}-\frac{6}{L^2} g_{\mu\nu}-\frac{1}{2} F_{\mu\rho}F_{\nu}^{\ \rho }
 +\frac{1}{8} F^2 g_{\mu\nu} \nonumber\\&-\frac{1}{2} F^{(5)}_{\mu\rho}F_{\nu}^{(5) \rho } +\frac{1}{8} F_{(5)}^{2}g_{\mu\nu}=0. \label{eom:Einstein}
\end{align}
The expectation values of the energy-momentum tensor and of the conserved currents in the field theory are obtained by the standard holographic prescription, i.e. varying the on-shell action with respect to the boundary value of the metric and gauge fields, respectively (see Refs. \cite{Zaanen:2015oix,Ammon:2015wua}). For the $U(1)_A\times U(1)_V$ model, we have 

\begin{align}
    &2\kappa^2\langle J^\alpha \rangle = \sqrt{-\gamma}n_\mu \left.\left( F^{\alpha\mu}+4\alpha \epsilon^{\mu\alpha\beta\gamma\delta}A_\beta F_{\gamma\delta} \right)\right|_{\partial \mathcal{M}} \label{vevJ}\\
    & 2\kappa^2\langle J_5^\alpha \rangle =\! \sqrt{-\gamma}n_\mu\!\! \left.\left( F_5^{\alpha\mu}\!+\!\frac{4}{3}\alpha \epsilon^{\mu\alpha\beta\gamma\delta}A_\beta F^5_{\gamma\delta} \right)\right|_{\partial \mathcal{M}}
    \label{vevJ5}\\
    & \kappa^2\langle T^{\alpha\beta} \rangle = \sqrt{-\gamma} \left.\left(-K^{\alpha \beta} + \gamma^{\alpha \beta} K \right)\right|_{\partial \mathcal{M}},
    \label{vevTmunu}
\end{align}
where $J$ and $J_5$ are the vector and axial currents, respectively. The boundary of spacetime is denoted $\partial \mathcal{M}$ and $n^\mu$ is the outward pointing unit vector orthonormal to $\partial \mathcal{M}$. The extrinsic curvature is denoted by $K$ and the induced metric on the boundary by $\gamma^{\alpha \beta}$.

In order to study the CME in this model, we need to construct an asymptotic AdS solution at finite temperature, axial charge and magnetic field. In the following, we explain how to implement each of them in two different setups: (a) a static, non-expanding plasma and (b) an expanding plasma. An arbitrary time-dependent magnetic field renders the system inhomogeneous and the resulting partial differential equations depend not only on the holographic coordinate and time, but also on the transverse plane coordinates, rendering the problem less tractable. In this work, we make the approximation that the magnetic field is small compared to the temperature of the system: $B/(\pi T)^2\ll 1$. This simplifying assumption effectively removes the dependence of the differential equation on the transverse plane coordinates to leading order. 

\begin{figure}[h]
    \centering
    \includegraphics[width=0.85\linewidth]{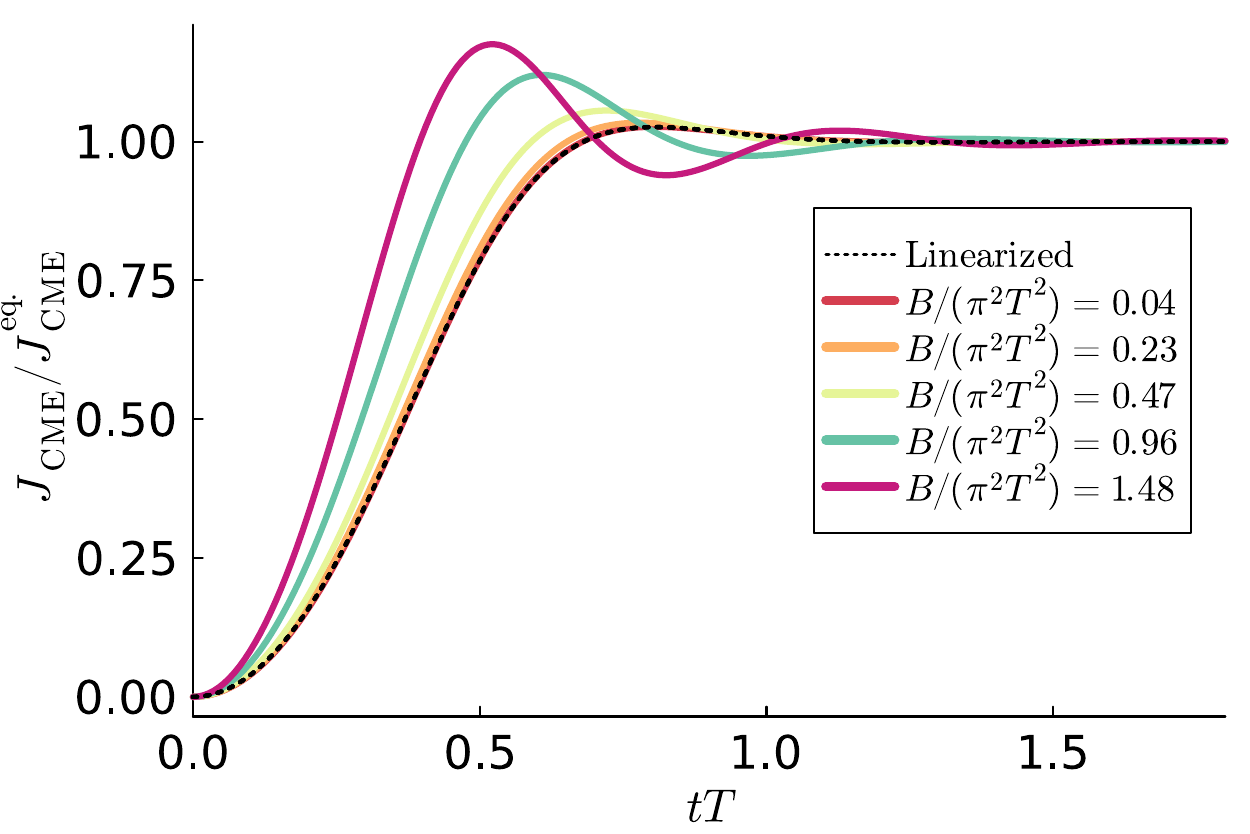}
    \caption{Ratio of the chiral magnetic current to its equilibrium value computed from the model with backreaction \cite{Ghosh:2021naw} (colored lines) and from the linearized equations of motion (black-dashed lines). The value of axial charge for each simulation is $2\kappa^2 n_5/T^3 = \{1.86,1.86,8.68,6.82,5.27\}$ corresponding to $B/(\pi T)^2 = \{0.04,0.23,0.47,0.96,1.48\}$ respectively.} 
    \label{fig:comparison}
\end{figure}

Indeed, for collisions at the RHIC accelerator with center-of-mass energy $\sqrt{s}=200$ GeV, we have a peak value of the magnetic field of the order of the pion mass $B \sim m_{\pi}^2\sim (140 \  \textrm{MeV})^2$ while the temperature is about $T\sim 300 \  \textrm{MeV}$, which gives $B/(\pi T)^2 \sim 2\times 10^{-2}\,$. The magnetic field quickly drops to even smaller values. A similar estimate for LHC energies with $B \sim 10 m_{\pi}^2$ and $T\sim 600-1000 \  \textrm{MeV}$ provides a maximum value for the ratio $B/(\pi T)^2 \sim (2-5)\times 10^{-2}\,$, where again this ratio drops as the magnetic field decays. Finally, a comparison between the results of the model at arbitrary (albeit constant) magnetic field \cite{Ghosh:2021naw} and the results treating the (constant) magnetic field as a perturbation \cite{PhysRevD.100.126024} are in qualitative agreement. Therefore, we construct a chirally imbalanced, finite temperature, (non)-expanding background and include the time dependent magnetic field as a perturbation on top of it. In Fig. \ref{fig:comparison}, we provide a comparison between the chiral magnetic current obtained for a non-expanding plasma at constant magnetic field with backreaction, as done in \cite{Ghosh:2021naw}, and the chiral magnetic current obtained from the linearized equations of motion at constant magnetic field for the same dimensionless ratios $B/T^2$ and $\mu_5/T$ in both cases. The details about the linearized equations and their solution are presented in Sec. \ref{sec:non_exp}. It is clear from Fig. \ref{fig:comparison} that the deviations between the linearized and backreacted solutions are no longer negligible if the magnetic field is larger than $B/(\pi T)^2\sim 0.5\,$. Consequently, the small magnetic field limit is a reasonable approximation for the holographic study of the QGP in heavy-ion collisions.

The finite temperature is attained in the holographic model by demanding that the solution features a black hole. Then the temperature of the dual field theory state is given by the temperature of such black hole. For the expanding case, the black hole is evolving with time, but an effective temperature can be defined as that of the apparent horizon. In the non-expanding case, the background is static and there is no ambiguity. As for the chiral imbalance, this can be introduced through the temporal component of the axial gauge field, which induces chiral charge via Eq. \eqref{vevJ5}. In the non-expanding case, a time-like killing vector $\eta^\nu$ can be used to define an axial chemical potential:
\begin{equation}
    \mu_5 = \int F_{\mu \nu}^5 \eta^\mu dx^\nu
\end{equation}
where the integral is performed from the boundary of spacetime up to the horizon.

\subsection{Non-expanding Plasma}\label{sec:non_exp}

\subsubsection{Background construction}

The background dual to a chirally imbalanced infinite non-expanding plasma is given by a chirally charged AdS black hole. The axial gauge field, in the radial gauge, and the ansatz for the metric in infalling Eddington-Finkelstein-like coordinates take the form
\begin{equation}\label{eq:axial_nonexp}
    A = \dfrac{1}{2}\,q_5 u^2 dv\,
\end{equation}
\begin{equation}\label{eq:metric_nonexp}
    ds^2 = -f(u)dv^2 - \dfrac{2 }{u^2} dv du + \dfrac{1}{u^2} (dx^2+dy^2+dz^2)\,
\end{equation}
where $u$ is the holographic radial coordinate, $v$ is the time-like coordinate, and $q_5$ is a constant controlling the chiral imbalance of the system. Note that the Eddington-Finkelstein time coincides with the time $t$ in Schwarzschild-like coordinates at the boundary ($u=0$).

\noindent
The blackening factor $f$ that solves the background equations of motion is

\begin{equation}\label{eq:f_nonexp}
    f(u) = \dfrac{1}{u^2}\left[1-\left(\dfrac{u}{u_h}\right)^4 - \dfrac{1}{12}\dfrac{u^4}{u_h^4}(q_5 u_h^3)^2\left(1-\dfrac{u^2}{u_h^2}\right)\right]
\end{equation}
The boundary is located at $u=0$ and the (outermost) horizon at $u_h\,$.

\subsubsection{Magnetizing the state}

We now introduce the time-dependent magnetic field to linear order in the perturbation parameter $\epsilon$. The ansatz for the vector gauge field, in the radial gauge, is given by  

\begin{equation}
    V = \epsilon \frac{1}{2} V_{\perp}(v,u)(y dx-x dy) + \epsilon V_z (v,u) dz\,.
\end{equation}
The field $V_{\perp}(v,u)$ contains information about a magnetic field $B$ pointing in the $z$ direction. In the case of a constant magnetic field, we have $V_{\perp}(v,u) = B$, whereas a time-dependent solution requires that $V_{\perp}$ depend on the two coordinates $v$ and $u$. The $z$ component of the vector gauge field captures the response of the chiral magnetic current to the presence of the magnetic field and chiral imbalance.

The background solution given in equations \eqref{eq:axial_nonexp} and \eqref{eq:metric_nonexp}  is not modified to linear order in $\epsilon$. The equations of motion obeyed by the $V_\perp$ and $V_z$ fields are of the form \ref{eom:Maxwell}:

\begin{align}\label{eq:linear-eoms1}
    &u\dot{V}'_{\perp}-\frac{1}{2} \dot{V}_{\perp} = \frac{1}{2}V_{\perp}'(u^2 f + u^3 f') +\frac{1}{2}  u^3 f V_{\perp}''  \\
    & u\dot{V}_z'-\frac{1}{2} \dot{V}_z = \frac{1}{2}V_z'(u^2 f + u^3 f') +\frac{1}{2}  u^3 f V_z'' + 4 q_5 \alpha u^3 V_{\perp}\label{eq:linear-eoms2}
\end{align}
The prime denotes derivatives with respect to the radial coordinate $u$, whereas the overdot is used for derivatives with respect to the time coordinate $v$. We can solve the previous equations asymptotically near the boundary. The result is
\begin{align}\label{eq:bdy_V_perp}
    V_{\perp} &= B_z + \dot{B}_z u + \left(V_{\perp}^{(2)}+\dfrac{1}{2} \ddot{B}_z\log u\right) u^2 +  \nonumber\\ &\left(\dot{V}_{\perp}^{(2)} -\frac{1}{3}\partial_v^3B_z + \frac{1}{2}\partial_v^3B_z \log u \right)u^3 + O(u^4)\,,
\end{align}
\begin{equation}\label{eq:bdy_V_z}
    V_z = V_z^{(2)} u^2 +  \dot{V}_{z}^{(2)} u^3 + \left(  \frac{5}{8} \ddot{V}_{z}^{(2)}  -\alpha q_5 B_z \right)u^4 + O(u^5)\,,
\end{equation}
where $B_z$ is the boundary magnetic field, while $V_\perp^{(2)}$ and $V_z^{(2)}$ are not fixed by the near-boundary analysis.
\subsubsection{Temperature and 1-point functions}

The temperature of the solution is that of the black hole, which is obtained as

\begin{equation}\label{temp_nonexp}
T = \dfrac{1}{2\pi}\left(-\frac{u_h^2}{2} f'(u_h)\right) = \dfrac{1}{u_h\pi}\left(1 - \dfrac{(q_5 u_h^3)}{24 }\right)\,,  
\end{equation}
where we have used the solution for the blackening function in the static plasma given in Eq.~\eqref{eq:f_nonexp}. The axial charge and chiral magnetic current, denoted as $n_5$ and $J_{\rm CME}$ respectively, are obtained substituting the near-boundary behavior of $A$ and $V$, given in Eqs. \eqref{eq:axial_nonexp}, \eqref{eq:bdy_V_perp} and \eqref{eq:bdy_V_z}, into the one-point functions \eqref{vevJ} and \eqref{vevJ5}, resulting into

\begin{align}\label{vevs:non_exp}
    n_5 &\equiv  \langle J^0_5 \rangle = \dfrac{q_5}{2\kappa^2}\,, \quad J_{\textrm{CME}} \equiv \langle J^z \rangle =  \dfrac{V_z^{(2)}}{\kappa^2}\,.
\end{align}

Note that solutions with a time-dependent magnetic field give rise to a non-vanishing value of $V_\perp^{(2)}$, which is ultimately related to the generation of currents in the transverse plane: $J^x$ and $J^y$. In this case, the on-shell action needs to be properly renormalized with additional counterterms that remove divergences proportional to the temporal gradients of the source term $\dot{B}_z$ and correct the finite contribution to Eq. \eqref{vevJ}. We do not pursue such renormalization, since our main interest resides on the chiral magnetic current, and instead define the transverse plane currents through the finite contribution of Eq. \eqref{vevJ} proportional to $V_\perp^{(2)}$:
\begin{equation}\label{eq:trans_nonexp}
    J^x \equiv  \frac{1}{4\kappa^2}y V_\perp^{(2)}\,,
     \quad  J^y \equiv -\frac{1}{4\kappa^2}x V_\perp^{(2)}\,.
\end{equation}
We further define $J_\perp$ as the curl of the circular current above:
\begin{equation}\label{eq:jprp}
    J_\perp\equiv\dfrac{1}{2\kappa^2}V_\perp^{(2)}\,.
\end{equation}
\subsubsection{Out-of-equilibrium initial state}

Starting from Eqs. \eqref{eq:linear-eoms1} and \eqref{eq:linear-eoms2}, we can first obtain the equilibrium configurations for the system. Setting the time derivatives of Eq. \eqref{eq:linear-eoms1} to zero, we integrate from the location of the horizon $u_h$ until an arbitrary point $u$ to obtain

\begin{equation}
    u f V_\perp '=0\,,
\end{equation}
which is solved by the constant $V_\perp=B_z$. Similarly, we can set the time derivatives of \eqref{eq:linear-eoms2} to zero, divide by $u^2$ and integrate once to obtain

\begin{equation}
    u f V_z' = -4q_5 \alpha (u^2-u_h^2) B_z\,.
\end{equation}
Evaluating the previous equation near the boundary $u\to 0$, and using the asymptotic expansion \eqref{eq:bdy_V_z}, reveals the well-known result that the time-independent configuration has a constant chiral magnetic current given by 
\begin{equation}
    2\kappa^2 J_{\textrm{CME}}^{\textrm{eq.}} = 8\alpha \mu_5 B_z\,.
\end{equation}
In our analysis, we will implement a time-dependent magnetic field $B(t)$. In this way, the out-of-equilibrium dynamics is driven by the gradients of $B$. We start with a state with zero chiral magnetic current. In addition, we start with some constant magnetic field $B_z(0)$ and provide the time derivative $\dot{B}_z$ as a boundary condition during the time evolution. Thus, any arbitrary profile $B(t)$ can be generated. In particular, the initial state is chosen as:

\begin{equation}
    V_z^{\textrm{ini.}} = 0 \qquad V_\perp^{\textrm{ini.}} = B_z(0) + \dot{B}_z(0)u\,.
\end{equation}
In section \ref{sec:RHIC}, we discuss the different magnetic field profiles $B_z(t)$ studied in this work.

\subsection{Expanding Plasma}\label{sec:expanding_plasma}

\subsubsection{Background construction}

Similarly to the previous case, we want a background that is dual to a chirally imbalanced thermal state. The magnetic field is later introduced as a perturbation on top of this background. The fact that the plasma is expanding can be realized by demanding that the conformal boundary undergoes Bjorken expansion at late times~\cite{Chesler:2009cy,Critelli:2018osu,Rougemont:2021qyk,Rougemont:2022piu,Grieninger:2022yps,Cartwright:2022hlg}. We choose that the expansion takes place along the $x$-direction, and express the boundary coordinates in terms of the rapidity $\eta$, the proper time $\tau$, and the coordinates transverse to the expansion $y$ and $z$. Specifically, we work with the ansatz for the gauge field and the metric given, in infalling Eddington-Finkelstein like coordinates, by
\begin{align}\label{eq:ansatz_expanding}
    A &= A_v dv\,, \nonumber\\
    ds^2 &= -f dv^2 -\dfrac{2}{u^2}dudv +\Sigma^2\left( e^{-2\xi}d\eta^2 + e^\xi dy^2+ e^\xi dz^2 \right)\,,
\end{align}
where ($A_v,f,\Sigma,\xi$) are functions of the Eddington-Finkelstein time $v$ and the holographic radial coordinate $u$.
The boundary is located at $u=0$, at which the Eddington-Finkelstein time $v$ coincides with the proper time $\tau$. Both time coordinates $\tau$ and $v$ will be used without distinction for boundary quantities. The chiral imbalance is implemented through the time component of the axial gauge field, whereas the thermal nature of the state is related to the presence of an apparent horizon at a distance $u_h(v)$. We demand that the metric functions near the boundary behave as 
\begin{equation}
    f\to \frac{1}{u^2}\,,\quad \Sigma \to \dfrac{\tau^{1/3}}{u}\,,\quad \xi\to -\frac{2}{3}\log\left(\tau\right)\,,
\end{equation}
 which is tantamount to demanding that the induced metric at the boundary asymptotes to 
 \begin{equation}\label{eq:exp_bdy_metric}
     ds^2\to \dfrac{1}{u^2}\left(-d\tau^2 + \tau^2 d\eta^2 + dy^2 + dz^2 \right)
 \end{equation}
where we can identify $\tau$ with the proper time and $\eta$ with the spacetime rapidity, which are related to the Minkowski time $t$ and spatial coordinate $x$ via
\begin{equation}
    t=\tau\cosh\eta\,,\qquad x = \tau\sinh\eta.
\end{equation}
The equations of motion for the axial gauge field \eqref{eom:axial} can be integrated once to give
\begin{equation}\label{eq:axial_gauge}
    A_v' = \dfrac{\tilde{q}_5}{u^2\Sigma^3}\,,
\end{equation}
where $\tilde{q}_5$ is an integration constant, which controls the axial charge $n_5$ of the system. The equations of motion for the metric tensor \eqref{eom:Einstein} can be compactly written as 

\begin{align}\label{eq:Eins_expanding}
    \left(u^2\Sigma'\right)' +\frac{1}{2}u^2\Sigma \left(\xi'\right)^2&=0\,, \nonumber\\
    \left(d\Sigma\right)'+2\dfrac{\Sigma'}{\Sigma}d\Sigma +\dfrac{2}{u^2}\Sigma - \dfrac{q_5^2}{12 u^2\Sigma^5} &= 0\,, \nonumber\\
     \left(d\xi\right)' + \dfrac{3\Sigma'}{2\Sigma}d\xi + \dfrac{3d\Sigma}{2\Sigma}\xi'& = 0\,, \nonumber\\ 
     f'' + \dfrac{2}{u}f' +\dfrac{12\Sigma'}{u^2\Sigma^2}d\Sigma -\dfrac{3\xi'}{u^2}d\xi +\dfrac{4}{u^2}-\dfrac{7q_5^2}{6u^4\Sigma^6}&=0\,, \nonumber\\
     dd\Sigma +\dfrac{1}{2}u^2 f'd\Sigma +\dfrac{1}{2}u^3f^2\Sigma'+\dfrac{1}{2}d\xi^2\Sigma +\dfrac{1}{8}u^4f^2\xi^2\Sigma&=0\,,
\end{align}

\noindent
where we have traded temporal derivatives with the derivatives along infalling null geodesics $d$:

\begin{equation}
    d = \partial_v - \dfrac{u^2 f }{2}\partial_u\ .
\end{equation}

\subsubsection{Late-time solution}\label{seclate}

We can construct an analytical solution for the late-time dynamics, in a similar spirit to the work of Ref. \cite{Janik:2005zt}, that is chirally imbalanced, expanding and boost invariant. This problem has been previously addressed in Ref. \cite{Kalaydzhyan:2010iv}. We repeat the derivation here for completeness and to set the notation. Inspired by the late-time expanding solution of the aforementioned work, we define the metric functions in Eq. \eqref{eq:ansatz_expanding} as

\begin{align}\label{eq:defs}
    &f(v,u) = \dfrac{F(v,u)}{u^2}\,,\qquad \Sigma(v,u) = \dfrac{v^{1/3}}{u} \sigma(v,u)\,,\nonumber\\&\xi(v,u) = -\dfrac{2}{3}\log\left(v\right)+\dots\,.
\end{align}
where the dots denote subleading contributions as $v\to \infty$.
Under these replacements the metric ansatz becomes

\begin{align}\label{eq:ansatz_expanding_late}
    ds^2 &= \dfrac{1}{u^2}\left(-F dv^2 -2 dudv +\sigma^2\left( v^2 d\eta^2 + dy^2+  dz^2 \right)\right)\,,
\end{align}
while the gauge field is still given by Eq. \eqref{eq:axial_gauge}, which under the definitions in Eq. \eqref{eq:defs} become

\begin{equation}\label{eq:Axprime}
    A'_v(v,u) = \dfrac{\tilde{q}_5}{v}\dfrac{u}{\sigma(v,u)^3}\,.
\end{equation}
We now expand the unkown functions $F$ and $\sigma$ at late times as

\begin{equation}\label{eq:asympt_late}
    F(v,u) = G\left(\zeta\right) + O(v^{-k})\,,\ \ \sigma(v,u) = S\left(\zeta\right)+ O(v^{-k})
\end{equation}
where $k$ is some positive number, such that the corrections in the previous expansion are negligible at late times ($v\to \infty$), and we defined the scaling variable\footnote{As discussed in Ref.~\cite{Janik:2005zt}, one can define the scaling coordinate $\zeta = u v^{-m/4}$, in which case the energy density decays as $\epsilon\sim \tau^{-m}$.} 
\begin{equation}\label{eqzeta}
    \zeta = \dfrac{u}{v^{1/3}}\,.
\end{equation}
The boundary is located at $\zeta=0$, while the bulk is defined for $\zeta>0$. In order to find the solution at late times, we substitute the definitions \eqref{eq:defs} and the asymptotic expansion \eqref{eq:asympt_late} into Einstein equations \eqref{eq:Eins_expanding}. Subsequently, we expand around $v\to \infty$ while keeping the scaling variable $\zeta$ fixed. To leading order, the first relation in Eqs. \eqref{eq:Eins_expanding} gives

\begin{equation}
    \partial_\zeta^2 S +\dots = 0 \quad \Rightarrow S = s_0 + s_1 \zeta + \dots
\end{equation}
where $s_0$ and $s_1$ are integration constants. The dots denote subleading contributions as $v\to \infty$.  We can set $s_0=1$ without loss of generality by simultaneous rescalings of the spatial coordinates $\eta,y,z$. The scaling variable is positive by definition $\zeta>0$, and therefore we have to choose $s_1\geq 0$ as well, otherwise the volume of the spatial sections, controlled by $S^2$, vanishes at some finite value of $\zeta$ and the geometry terminates at that point. We choose $s_1=0$ for simplicity. Then, we find that axial gauge field obtained from \eqref{eq:Axprime} is

\begin{equation}\label{axialgaugelate}
    A_v(v,u) = \dfrac{1}{2}\dfrac{\tilde{q}_5}{v} u^2 +\dots \,.
\end{equation}

Substituting the previous solutions into Einstein equations \eqref{eq:Eins_expanding} at late times gives two differential equations for the function $G$:

\begin{align}
    &\frac{\zeta  }{2}\partial_\zeta G -2 G-\frac{\tilde{q}_5^2 \zeta^6}{12}+2 = 0\\
   &\dfrac{1}{2}\zeta^2\partial_\zeta^2 G -3 \zeta\partial_\zeta G + 6G -  6-\frac{\tilde{q}_5^2 \zeta^6}{4}=0
\end{align}
where we have omitted subleading contributions at late times.
The system of equations is consistent, in the sense that any solution to the first order differential equation also satisfies the second order differential equation. In fact, the equations can be solved exactly, giving

\begin{equation}\label{eq:Gz}
    G(\zeta) = 1 + g_4 \zeta^4 + \dfrac{\tilde{q}_5^2}{12}\zeta^6
\end{equation}
with $g_4$ an integration constant that controls the energy density of the solution. Rewriting the solution in terms of the $v,u$ coordinates we obtain
\begin{align}\label{eq:expanding_late}
    ds^2 &= \dfrac{1}{u^2}\left[- \left( 1 + \frac{g_4}{v^{4/3}} u^4 + \frac{1}{12}\dfrac{\tilde{q}_5^2}{v^2}u^6\right)dv^2 -2 dudv \right.\nonumber\\&\left.+\left( v^2 d\eta^2 + dy^2+  dz^2 \right)\right]\,,
\end{align}
which is somewhat similar to a Reissner-Nordstr\"om AdS black hole except that in Eq. \eqref{eq:expanding_late} both the axial charge and energy density decay with time. Accordingly, the effective temperature and chemical potential of the dual field theory decay with time as
\begin{align}\label{eq:temp_exp}
    T\simeq &\dfrac{1}{2\pi u_h}\left(1 -\dfrac{g_4}{\tau^{4/3}}u_h^4 -\frac{1}{6}\dfrac{\tilde{q}_5^2}{\tau^2}u_h^6\right) \nonumber\\=& \dfrac{1}{\tau^{1/3}}\dfrac{1}{\pi \zeta_h}\left(1-\frac{1}{24}\tilde{q}_5^2 \zeta_h^6\right)\,,\\
   \mu_5 \simeq  &  \dfrac{q_5}{2\tau}u_h^2\,,
\end{align}
where $\zeta_h$ is a constant corresponding to the smallest positive zero of $G(\zeta)$ [c.f. Eq. \eqref{eq:Gz}]. Note that the location of the horizon in infalling Eddington-Finkelstein like coordinates changes with time according to 
\begin{equation}\label{u_h(v)}
    u_h(v) = \zeta_h v^{1/3}\,.
\end{equation}
We will come back to this observation in Sec. \ref{sec:numsetup}. 

\begin{figure}
    \centering
    \begin{tabular}{c}
    \includegraphics[width=0.85\linewidth]{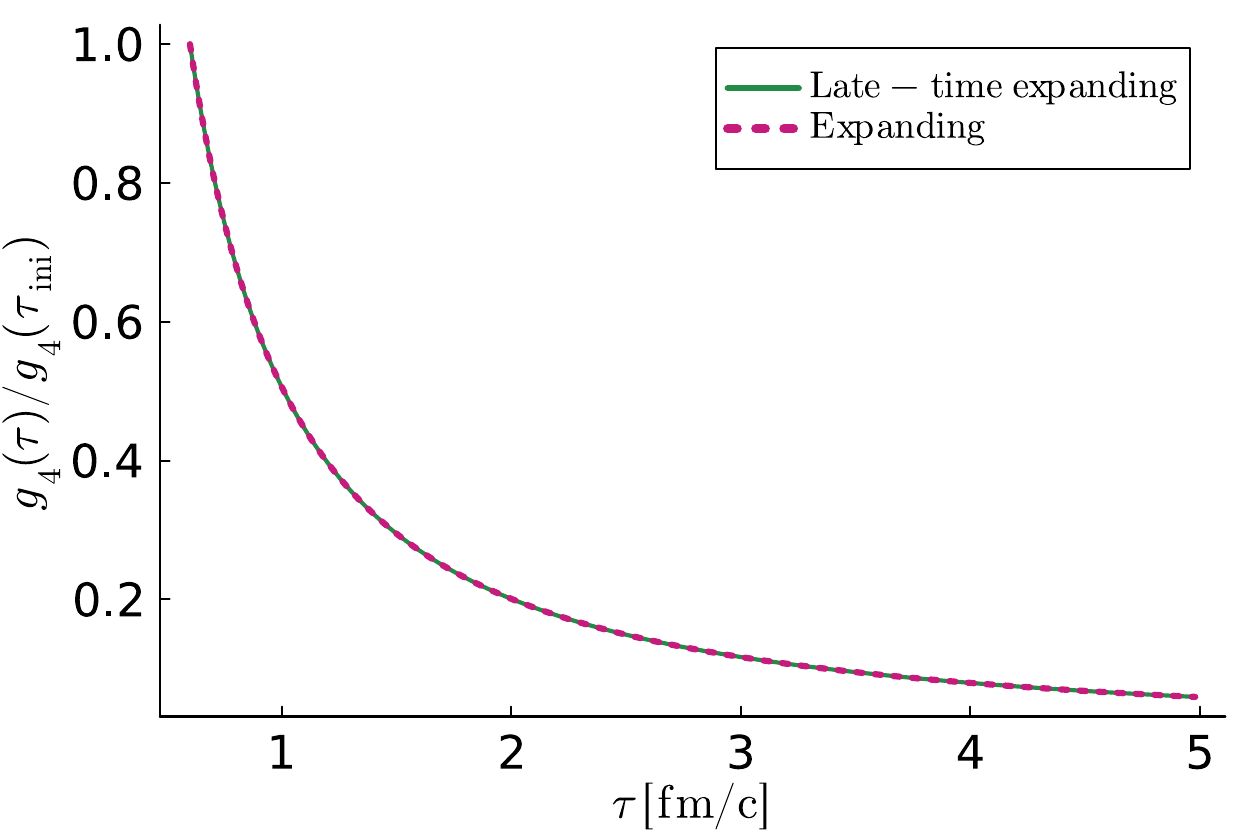} \\ \includegraphics[width=0.85\linewidth]{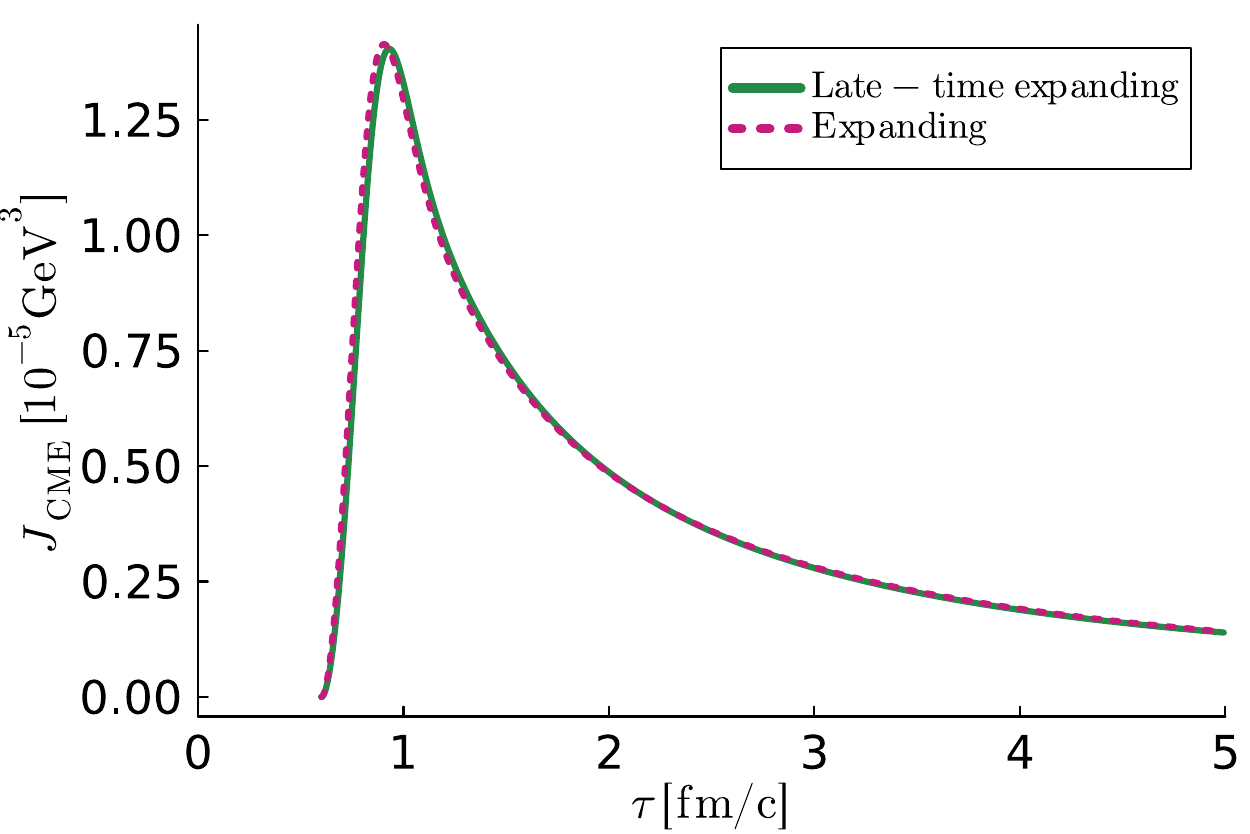}
    \end{tabular}
    \caption{(Top panel) Time dependent integration constant for the blackening function $f$ controlling the energy density, normalized to its initial value. In the late-time expanding solution \eqref{flate}, the shown quantity is given by $g_4(\tau) = g_4/\tau^{4/3}$. (Bottom panel) Chiral magnetic current. (Both) Comparison between the solution of Eqs. \eqref{eq:Eins_expanding} and the late-time approximation of Eq. \eqref{eq:expanding_late} for the same thermodynamic parameters and initial time $\tau_{\rm ini}=0.6$ fm/c. }
    \label{fig:comparison-latetime}
\end{figure}

The agreement between the late-time solution compared to the full expanding solution to Eqs. \eqref{eq:Eins_expanding} is presented in Fig. \ref{fig:comparison-latetime}. The energy density, obtained from Eqs. \eqref{eq:Eins_expanding} and controlled by $g_4(v)$,  quickly follows the Bjorken scaling $\epsilon\sim \tau^{-4/3}$. The chiral magnetic current, shown in the bottom panel, is also similar in both approaches, with a slight deviation at early times. We will make use of the late-time solution for our simulations.

\subsubsection{Magnetizing the state}

We assume that the magnetic field is small and write the following ansatz for the vector gauge field:

\begin{equation}
    V = \epsilon\ \eta V_\perp(v,u) dy +\epsilon\ V_z(v,u) dz\,.
\end{equation}
The interpretation of $V_\perp$ and $V_z$ is similar to the non-expanding case. The former controls the magnetic field pointing in the $z$ direction while the latter encodes the chiral magnetic current generated in the system. The equations of motion \eqref{eom:Maxwell}, to leading order in $\epsilon$, become 
\begin{align}
    & \left(\partial_u -\frac{1}{2}\xi' + \dfrac{1}{2}\dfrac{\Sigma'}{\Sigma}\right)dV_\perp - \left(\frac{1}{2}d\xi - \dfrac{d\Sigma}{2\Sigma}\right)V_\perp'=0\,,\\
    & \left(\partial_u -\frac{1}{2}\xi' + \dfrac{1}{2}\dfrac{\Sigma'}{\Sigma}\right)dV_z - \left(\frac{1}{2}d\xi - \dfrac{d\Sigma}{2\Sigma}\right)V_z' =- \dfrac{4\alpha \tilde{q}_5 V_\perp e^\xi}{u^2\Sigma^4} \,.
\end{align}
In the late-time expanding ansatz \eqref{eq:ansatz_expanding_late}, the equations further simplify to 
\begin{align}\label{eq:gaugelate}
    & \left(\partial_u -\dfrac{1}{2u}\right)dV_\perp + \left(\dfrac{1}{2v}+\dfrac{u}{4}f\right)V_\perp'=0\,,\\
    & \left(\partial_u -\dfrac{1}{2u}\right)dV_z + \left(\dfrac{1}{2v}+\dfrac{u}{4}f\right)V_z' =- \dfrac{4\alpha \tilde{q}_5 u^2 V_\perp }{v^2} \,,\label{eq:gaugelate2}
\end{align}
where $f$ is given by
\begin{equation}\label{flate}
    f = \dfrac{1}{u^2}\left( 1 + \frac{g_4}{v^{4/3}} u^4 + \frac{1}{12}\dfrac{\tilde{q}_5^2}{v^2}u^6\right)\,.
\end{equation}
Near the boundary ($u=0$) the gauge field components $V_z$ and $V_\perp$ that solve Eq \eqref{eq:gaugelate} behave as
\begin{equation}\label{bdyVtransexp}
    V_\perp = b_z + \dot{b}_z u + \left(b_\perp^{(2)} + \dfrac{\partial_v(v \dot{b}_z)}{2v}\log u\right)u^2 + O(u^3)\,,
\end{equation}
\begin{equation}\label{bdyVzexp}
    V_z = V_z^{(2)} u^2 + \left( \dfrac{2V_z^{(2)}}{3v} + \dot{V}_z^{(2)} \right)u^3+O(u^4) \,,
\end{equation}
where we have set the leading solution of the latter to zero, $V_z^{(0)}=0$, which is tantamount to not sourcing the chiral magnetic current in the dual field theory, while $V_z^{(2)}$, $V_\perp^{(2)}$ and $b_z$ are integration constants. As we shall soon see, the integration constants control the CME response, the transverse current, and the magnetic field, respectively. As pointed out in \cite{Cartwright:2021maz}, the magnetic field in the co-moving frame $u^{\mu} = (1,0,0,0)$ gets corrected by a factor of $\tau^{-1}$:
\begin{equation}
 B^z  =\dfrac{1}{2}\epsilon^{z\nu\rho\sigma}u_\nu \mathbf{F}_{\rho\sigma} = \dfrac{b_z(\tau)}{\tau}\,,   
\end{equation}
where $\mathbf{F}_{\rho\sigma}$ is the field strength of the boundary field theory, obtained from the corresponding gauge field
\begin{equation}
    \mathbf{V} = \lim_{u\to0}V = \eta b_zdy\,. 
\end{equation}

The axial charge and chiral magnetic current are, in this case
\begin{align}\label{vevs_expanding}
    n_5 &= \langle J_5^0\rangle =\dfrac{1}{2\kappa^2}\dfrac{\tilde{q}_5}{\tau}\,,\quad J_{\textrm{CME}} = \langle J^z\rangle = \dfrac{1}{\kappa^2}V_z^{(2)}\,.
\end{align}

\subsubsection{Initial state and late-time response}\label{sec:inilate}

Similarly to the non-expanding case, we consider an initial state with $V_z=0$, thus with vanishing chiral magnetic current, and a radial profile for $V_\perp$ that contains information about the magnetic field at initial time as well as its first time derivative:
\begin{equation}
    V_z^{\textrm{ini.}} = 0\,,\quad V_\perp^{ini} = b_z(0)+\dot{b}_z(0) u\,.
\end{equation}

We can also extract the expected late-time behavior of the chiral magnetic current following a similar strategy used to obtain the late-time geometry in Sec. \ref{seclate}. In particular, we consider the following late-time ansatz for the components of the vector gauge field,
\begin{equation}\label{lateans}
    V_\perp (v,u)= \tilde{V}_\perp(\zeta)b_z(v) + \dots\,,\quad V_z(v,u)= \tilde{V}_z(\zeta)g(v)+\dots\,,
\end{equation}
where $\zeta$ is the scaling coordinate defined in Eq. \eqref{eqzeta}. The function $b_z(v)$ is arbitrary and eventually related to the magnetic field profile, while the function $g(v)$ will be determined from the equations of motion. The dots denote subleading corrections as $v\to \infty$. Then, the equations of motion \eqref{eq:gaugelate} can be written as
\begin{align}\label{lateb}
      &v^{2/3}b_z(v)  \left[\partial_\zeta\tilde{V}_\perp (\zeta\partial_z-1)G + \zeta G\partial_\zeta^2\tilde{V}_\perp\right]\nonumber\\& +  v \dot{b}_z(v) \left(1-2\zeta \partial_\zeta\right)\tilde{V}_\perp +\dots=0\,,
\end{align}
\begin{align}\label{latecme}
    &  v^{2/3}g(v)  \left[\partial_\zeta\tilde{V}_z (\zeta\partial_\zeta-1)G + \zeta G\partial_\zeta^2\tilde{V}_z\right]\nonumber\\&+v \dot{g}(v) \left(1-2\zeta \partial_\zeta\right)\tilde{V}_z +8\tilde{q}_5\alpha\zeta^3\tilde{V}_\perp b_z(v) + \dots=0\,.
\end{align}
where $G$ is defined in Eq. \eqref{eq:Gz}. The second term in Eq. \eqref{lateb} is subleading in the late-time expansion for a large class of choices of the magnetic field $b_z(v)$, all those for which $v^{2/3}|b_z|>v |\dot{b}_z|$, in particular for any magnetic field that decays with a power-like behavior. At the end of this section, we comment on the exceptions to the previous inequality. Neglecting the second term in Eq. \eqref{lateb}, it is easy to show that the only solution that is regular at the horizon $\zeta_h$, where $G(\zeta_h)=0$, is $V_\perp =$constant, where the constant can be set to unity without loss of generality, since it can be absorbed into $b_z$. Regarding Eq. \eqref{latecme}, we can similarly neglect the term proportional to $\dot{g}(v)$. In addition, a non-trivial solution is attained when both $v^{2/3} g(v)$ and $b_z(v)$ are of the same order at late times. Therefore, $g(v) = b_z(v)v^{-2/3}$ up to an irrelevant constant that can be reabsorbed into $\tilde{V}_z$. Under these considerations, we can integrate Eq. \eqref{latecme} once to find
\begin{equation}\label{eq:solvzp}
    \partial_\zeta\tilde{V}_z = -\dfrac{4\alpha\tilde{q}_5\zeta(\zeta^2-\zeta_h^2)}{G}\,,
\end{equation}
where we have set the integration constant such that the solution is regular at the location of the horizon $\zeta_h$. We can now substitute $G(\zeta)$ from Eq. \eqref{eq:Gz} and solve Eq. \eqref{eq:solvzp} near the boundary ($\zeta=0$) to obtain
\begin{equation}
    \tilde{V}_z = 2\alpha \tilde{q}_5\zeta_h^2\zeta^2 + O(\zeta^3)\,.
\end{equation}
Finally, we can go back to the original coordinates $u$ and $v$ and calculate the asymptotic form of $V_z(v,u)$ of Eq. \eqref{lateans},
\begin{equation}
    V_z = \tilde{V}_z g(v) +\dots= 4\alpha \left(\dfrac{1}{2}\tilde{q}_5\zeta_h^2\right) \dfrac{b_z(v)}{v^{4/3}}u^2 + O(u^3)+\dots\,,
\end{equation}
from which we obtain that the chiral magnetic current at late times follows the magnetic field decay $b_z(\tau)/\tau$ times a factor of $\tau^{-1/3}$. In particular, for the purely Bjorken expansion $b_z=1$, the chiral magnetic current decays as $\tau^{4/3}$, as it was numerically verified in Ref. \cite{Grieninger:2023myf}. As a last step, we can define the effective chemical potential in the late-time expanding solution from the temporal component of the axial gauge field \eqref{axialgaugelate},
\begin{equation}\label{mu5exp}
    \mu_5 = \dfrac{1}{2}\dfrac{\tilde{q}_5u_h^2}{\tau} = \dfrac{1}{2}\dfrac{\tilde{q}_5\zeta_h^2}{\tau^{1/3}}\,.
\end{equation}
Then, the chiral magnetic current can be compactly written as 
\begin{equation}
    2\kappa^2 J^{\text{ quasi-eq}}_{\textrm{CME}} = 8\alpha\mu_5 \dfrac{b_z(\tau)}{\tau}=8\alpha\mu_5 B_z\,,
\end{equation}
recovering the (quasi)-equilibrium formula for the expanding case at late times. A numerical verification of the previous formula is presented in Sec. \ref{sec:RHIC}. 

Notably, the assumption that $v^{2/3}|b_z|>v |\dot{b}_z|$, made in Eqs. \eqref{lateb} and \eqref{latecme}, does not hold for the cases where the magnetic field decays exponentially at late times. In general, if the late time behavior of the magnetic fields is such that $v^{2/3}|b_z|<v |\dot{b}_z|$, then one should neglect the first term in Eqs. \eqref{lateb} and \eqref{latecme}. In such a case, Eq. \eqref{lateb} becomes
\begin{equation}
    (1-2\zeta\partial_\zeta)\tilde{V}_\perp = 0 \Rightarrow \tilde{V}_\perp=c_0 \sqrt{\zeta}\,.
\end{equation}
Such a solution is incompatible with the asymptotic behavior of $V_\perp$ near the boundary in Eq. \eqref{bdyVtransexp}. Therefore, we are forced to set the integration constant $c_0=0$. Similarly, form Eq. \eqref{latecme} we find that $\tilde{V}_z = c_1 \sqrt{\zeta}$ and we set $c_1=0$ for consistency with the near-boundary asymptotics of Eq. \eqref{bdyVzexp}. The previous result can be interpreted as the fact that if the magnetic field decays sufficiently fast, i.e. $v  |\dot b_z|<v^{2/3}|b_z|$, then it is incompatible with a Bjorken-like regime. In other words, such a magnetic field would be negligible while the plasma continues to expand. As a result, the chiral magnetic current similarly vanishes at late times for such rapidly vanishing magnetic fields.

\section{Thermodynamic parameters and Numerical Setup}\label{sec:III}

In this section, we discuss the physical parameters corresponding to the quark-gluon plasma generated in heavy-ion collisions as well as the mapping to the values used in the numerical simulations.

\subsection{Thermodynamic parameters}\label{sec:thermo}

The model presented in Sec.~\ref{sec:model} allows us to study the dynamical response of the chiral magnetic current in an infinite (non)-expanding plasma, characterized by the energy density $\epsilon$, equivalently the temperature $T$, the axial charge $n_5$, and the time-dependent magnetic field $B$. For the case of an expanding plasma, there is an extra parameter related to the initial time of the simulation $\tau_{\rm ini.}\neq0$. In this section, we discuss the parameters for the quark-gluon plasma generated in heavy ion collisions, as well as their scaling with the center-of-mass energy of the collision $\sqrt{s}$, that we use in the numerical studies of Sec. \ref{sec:results}. We scale all quantities with the collision energy. 

\subsubsection{Temperature and axial charge}

The temperature of the QGP can be extracted from the leptons emitted in the collision. Although the effective temperature extracted from photon production is relatively insensitive to the actual temperature of the plasma, the effective temperature obtained from the dilepton spectra can be used as a sensitive probe of the true temperature of the plasma \cite{Massen:2024pnj}, since the leptons have a longer mean free path.  A recent study from the STAR collaboration \cite{STAR:2024bpc} concludes that the effective temperature of the QGP at energies $\sqrt{s} = 27$ GeV and $\sqrt{s}=54.4$ GeV in Au-Au collisions extracted from the low mass region thermal dilepton spectra is about $167\pm21\pm18$ MeV and $172\pm13\pm18$ MeV respectively. A similar extraction from the intermediate mass region, corresponding to dileptons emitted at an earlier stage of the plasma, reveals $T_{\rm eff.} = 280\pm64\pm10$ MeV and $303\pm59\pm28$ MeV respectively. The quoted values of temperatures for the low and intermediate mass regions are compatible with those used in the holographic study  \cite{Cartwright:2021maz} (c.f. Table I of Ref. \cite{Cartwright:2021maz}). In this paper, we fit the values of Ref. \cite{Cartwright:2021maz} and extrapolate to different energies. Specifically, our obtained fit is 
\begin{equation}\label{T_energy}
    T(\sqrt{s}) = 96.50(\sqrt{s})^{0.23}-23.51\ \textrm{MeV}\,,
\end{equation}
where the center of mass energy is measured in GeV. Although the previous scaling does not follow from physical considerations, it is sufficient to set the value of the temperature at the collision energies considered in Sec. \ref{sec:results}: $\sqrt{s} = 19-400$ GeV.

The chiral imbalance of the plasma, controlled by the axial charge $n_5$, is similarly affected by uncertainties. In this work, we assume that the axial charge scales with energy as $(\sqrt{s})^{1/3}$ and use the value at $\sqrt{s}=200$ GeV as a benchmark. The estimate given in Ref. \cite{Cartwright:2021maz}, supported by references therein, is $n_5(200\ {\rm GeV}) = 0.0027 \ {\rm GeV}^3$. Note that there are different estimates of the amount of axial charge, for example Ref. \cite{Shi:2017cpu}, that suggest $n_5/\mathbf{s} = 0.065$ at $200$ GeV, where $\mathbf{s}$ is the entropy density of the plasma. This value is approximately $10$ times larger than the one considered in \cite{Cartwright:2021maz}. However, in both cases the axial charge is sufficiently small compared to the temperature of the plasma and therefore the difference between using either of the values is expected to be an overall multiplicative factor, as we already argued in Ref. \cite{Ghosh:2021naw}. All in all, we set 
\begin{equation}\label{n5_energy}
    n_5(\sqrt{s}) = 0.0027 \left(\dfrac{\sqrt{s}}{200\ {\rm GeV}}\right)^{1/3}. 
\end{equation}

In the non-expanding case, both temperature $T$ and axial charge $n_5$ are constant in time, and we employ Eqs. \eqref{T_energy} and \eqref{n5_energy} to set the values of temperature and axial charge in the plasma. For the expanding case, the two quantities decay with time and we use Eqs. \eqref{T_energy} and \eqref{n5_energy} to set their respective values at equilibration time, i.e. $n_5^0 = n_5(\tau_0)$ and $T_0 = T(\tau_0)$ as given in Eqs. \eqref{eq:temp_exp} and \eqref{vevs_expanding}. We take the equilibration time to be $\tau_0 = 0.6$ fm/c.

\subsubsection{Magnetic field}

We now turn our attention to the main focus of this work: the inclusion of a time-dependent magnetic field. The peak value of the magnetic field $B_{\rm \max.}$ is reached at the time when the colliding nuclei have maximum overlap, which we set to $t_*=\tau_*=0\,$. The overall magnitude of the magnetic field can be estimated to scale as $B_{\rm max.}\sim\sqrt{s}$, see Ref. \cite{Mayer:2024kkv} for a more extensive discussion. We set the value of the magnetic field at $\sqrt{s}=200$ GeV to be $B_{\rm max.} = m_\pi^2$, with $m_\pi = 140$ MeV the pion mass. Therefore, we work with 
\begin{equation}\label{B_energy}
    B_{\rm max.}(\sqrt{s}) = \left(\dfrac{\sqrt{s}}{200\ \rm{GeV}}\right)m_\pi^2
\end{equation}

There are several works in the literature that address the question of the time dependence of the magnetic field. In this paper, we consider several of them, which we introduce in the following, in order to extract more robust conclusions. The profiles presented will be considered both in the expanding and non-expanding backgrounds. For compactness, we express the magnetic field as a function of $\tau$, while it is understood that in the non-expanding scenario one replaces $\tau\to t$.

We first consider the simplest profile, where the magnetic field follows the Bjorken expanding scaling
\begin{equation}\label{profileA}
   \textbf{Profile A:} \qquad B(\tau) = \dfrac{B_{\textrm{max.}}\tau_{\rm ini.}}{\tau}\,.
\end{equation}
Such a profile is obtained in the expanding plasma of Sec. \ref{sec:expanding_plasma} with a time-independent $V_\perp$. This profile was considered in Ref. \cite{Cartwright:2021maz} and is included here for comparison purposes. The profile \ref{profileA} is divergent as $\tau\to0$ and the estimation of the magnetic field given in Eq. \eqref{B_energy} is set at the initial time of the simulation $\tau_{\rm ini.}$.

In Ref. \cite{Guo:2019joy}, three different temporal dependencies of the magnetic field were considered that would explain the observed polarization measurements in the $\Lambda$ and $\overline{\Lambda}$ baryons. Each of them contains a parameter $\tau_B$ controlling the lifetime of the magnetic field and its scaling with energy. In particular, they introduce the following time-evolving magnetic fields:
\begin{multline}\label{ProfileB}
    \textbf{Profile B:}\\ \quad B = \dfrac{B_{\rm max.}}{1+\left(\tau/\tau_B\right)^2}\,,\quad \tau_B = \dfrac{92 \,{\rm GeV}\cdot{\rm fm/c}}{\sqrt{s}}\,.
\end{multline}
\begin{multline}
    \textbf{Profile C:}\\ \hspace{-0.3cm} \quad B = \dfrac{B_{\rm max.}}{\left[1+\left(\tau/\tau_B\right)^2\right]^{3/2}}\,,\ \ \tau_B = \dfrac{125 \,{\rm GeV}\cdot{\rm fm/c}}{\sqrt{s}}\,.
\end{multline}
\begin{multline}\label{profileD}
    \textbf{Profile D:}\\ \quad B = B_{\rm max.} e^{-\tau/\tau_B}\,,\quad \tau_B = \dfrac{128 \,{\rm GeV}\cdot{\rm fm/c}}{\sqrt{s}}\,.
\end{multline}

Another important profile that we consider is that of Ref. \cite{Mayer:2024kkv} (c.f. blue curve of Fig. 5 in  Ref. \cite{Mayer:2024kkv}). The evolution of the magnitude of the magnetic field is obtained from magnetohydrodynamic simulations of Au-Au collisions at $\sqrt{s}=200$ GeV. We shall refer to the aforementioned profile as
\begin{equation}
    \textbf{BHAC-QGP:}\quad B(\tau) \ \text{from Ref. \cite{Mayer:2024kkv}}\,. 
\end{equation}
In order to extrapolate the BHAC-QGP profile to different collision energies, we fit the magnetic field to a function of the form $1/\sinh(\tau/\tau_B)$, which follows a Bjorken-like evolution at early times and decays exponentially at late times. The fitting parameter is found to be $\tau_B =2.42$ fm/c. The lifetime of the magnetic field in this case is controlled by $\tau_B$, which we assume scales as the parameters of profiles B, C and D, namely $\tau_B\sim (\sqrt{s})^{-1}$. Consequently, we consider the following profile to extrapolate the BHAC-QGP data to different collision energies:
\begin{multline}\label{profileE}
    \textbf{Profile E:} \\ B = B_{\rm max.}\dfrac{\sinh(\tau_{\rm ini.}/\tau_B)}{\sinh(\tau/\tau_B)}\,,\quad\tau_B = \dfrac{484 \,{\rm GeV}\cdot{\rm fm/c}}{\sqrt{s}}\,,
\end{multline}
where the value of $\tau_B$ directly follows from the fit to the BHAC-QGP data. The late-time exponential behavior of profile E is suggestive of profile D. For this reason, we also consider profile $\overline{\rm E}$, which has the same functional dependence as in Eq. \eqref{profileE} but with the lifetime of Eq. \eqref{profileD}. 

\begin{figure}
    \centering
    \includegraphics[width=0.95\linewidth]{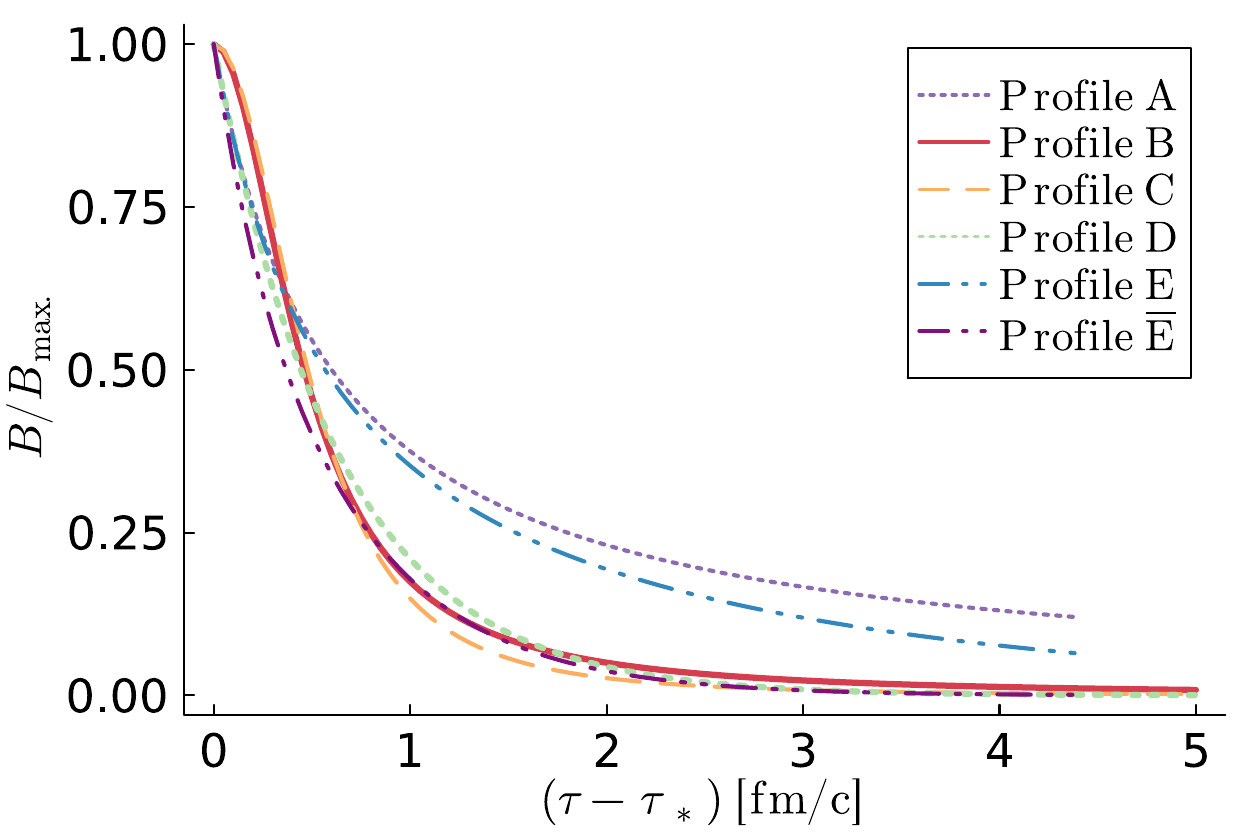}
    \caption{Magnetic fields described in Sec. \ref{sec:thermo} normalized to $B_{\rm \max.}$ for the center-of-mass energy $\sqrt{s}=200$ GeV. The profiles A, E and $\overline{\rm E}$ are initialized at the equilibration time $\tau_*=\tau_0 =0.6$ fm/c, while the profiles B, C and D are initialized at the time of maximum overlap $\tau_* =0\,$. }
    \label{fig:magneticfields}
\end{figure}

All profiles considered in this paper are shown in Fig. \ref{fig:magneticfields} at a center-of-mass energy of $\sqrt{s}=200$ GeV. The BHAC-QGP magnetic field overlaps with the fit that leads to profile E.  The profiles B, C and D were obtained by matching to the same physical phenomenon and are qualitatively similar to each other. In addition, they are systematically smaller than the other profiles considered in this work. Note that the hierarchy of magnetic fields, that is which magnetic field is bigger (smaller), changes at different collision energies due to the scaling of the lifetimes $\tau_B$. The crossing of hierarchies turns out to be crucial for the interpretation of the results in Sec. \ref{sec:results}.

\subsubsection{Initial time}

In the expanding scenario of Sec. \ref{sec:expanding_plasma}, there is an extra parameter that necessarily enters in the description of the system: the initial time of the simulation. In this respect, we follow the considerations of Ref. \cite{Huang:2022qdn}, for which the maximum overlap of the nuclei in the collision takes place at $t=\tau=0$. In the range between $\tau\in(0,0.1)$ fm/c, the plasma is assumed to be gluon dominated and therefore there is no response from the quark degrees of freedom, consequently no chiral magnetic current develops. Shortly after, for $\tau\in(0.1,\tau_0)$ fm/c, the system is out-of-equilibrium but both quarks and anti-quarks are present in the plasma. At this stage, the plasma is Bjorken expanding, which we describe with the metric background in Eq. \eqref{eq:expanding_late}. The parameter $\tau_0$ is the equilibration time, which we take to be $\tau_0=0.6$ fm/c. After the temperature drops to the critical value, the plasma hadronizes and the holographic description given by model \eqref{Bt:action} is no longer applicable. 

The previous discussion leads us to establish the initial time of the simulation as $\tau_{\rm ini.} = 0.1$ fm/c. Such an early time can be argued to be in conflict with the purely Bjorken expanding background described through the dual metric \eqref{eq:expanding_late}, and we further consider $\tau_{\rm ini.}=0.6$ fm/c, for which we have shown in Fig. \ref{fig:comparison-latetime} that the evolution in the background \eqref{eq:expanding_late} is expected to be a sufficiently good approximation.  

\subsection{Numerical setup}\label{sec:numsetup}

The dynamical response of the chiral magnetic current to the time-dependent magnetic field is encoded in the near boundary expansion of the holographic gauge field $V^z$. In order to access it, we solve the differential equations \eqref{eq:linear-eoms1} and \eqref{eq:linear-eoms2} for the non-expanding and \eqref{eq:gaugelate} and \eqref{eq:gaugelate2} for the expanding case respectively, with the initial state described in Sec. \ref{sec:model}. As it is standard, we solve the differential equations with pseudo-spectral methods in the radial direction while we implement a Runge-Kutta routine for the temporal evolution. Further details can be found in \cite{Ghosh:2021naw,Cartwright:2021maz,Grieninger:2023myf} and references therein. 

Note that the model (Sec.~\ref{sec:model}) exhibits a scaling symmetry, $ (v,u,x,y,z)\to\lambda (v,u,x,y,z)$, with the thermodynamic parameters transforming according to their scaling dimension. Thus, it is enough to consider the dimensionless combinations of the parameters given in Sec. \ref{sec:thermo}. The dimensionless combinations fix all parameters of the model except one, which we take to be the location of the horizon $u_h$ in the non-expanding case, and the integration constant $g_4$ [see Eq. \eqref{flate}] in the expanding case. In the non-expanding case, we arbitrarily set $u_h=1$ and perform the radial integration over the region between the boundary and the horizon, i.e. $u\in[0,1]$.

For the expanding case, the situation is slightly different. On the one hand, the implementation of the pseudo-spectral methods require that the radial domain of integration be rectangular, that is, a fixed interval for all times. On the other hand, the integration needs to be performed at least over the region causally connected to the boundary, in other words, between the boundary at $u=0$ and the horizon at $u_h$. However, in the expanding case, the location of the horizon changes with time according to Eq. \eqref{u_h(v)}. In order to have a rectangular domain that always includes the causally connected region to the boundary we integrate over $u\in[0,u_h(v_{\rm end.})]$, where $u_h(v_{\rm end.})$ is the location of the horizon at the final time of the simulation $v_{\rm end.}$. Finally, we use the integration constant $g_4$ such that the final location of the horizon is $u_h(v_{\rm end.}) =1\,$, and the integration domain is $u\in[0,1]$ also in the expanding case. 

\section{Numerical results}\label{sec:results}

We now combine the physical considerations of Sec. \ref{sec:III} with the holographic model of Sec. \ref{sec:model} in both an expanding and non-expanding plasmas under the approximation that the magnetic field is sufficiently small. Firstly, in Subsec. \ref{sec:RHIC} we discuss the sensitivity of the CME signal to the different magnetic fields displayed in Fig. \ref{fig:magneticfields}. Then, in Subsec. \ref{sec:energies}, we make use of the scaling with the energy of the collision $\sqrt{s}$ to draw conclusions of the CME signal as a function of energy, in a similar spirit as in Ref. \cite{Cartwright:2021maz}.

\subsection{Sensitivity at $\sqrt{s} = 200$ GeV}\label{sec:RHIC}

\begin{figure}
    \centering
    \begin{tabular}{c}
    \includegraphics[width=0.85\linewidth]{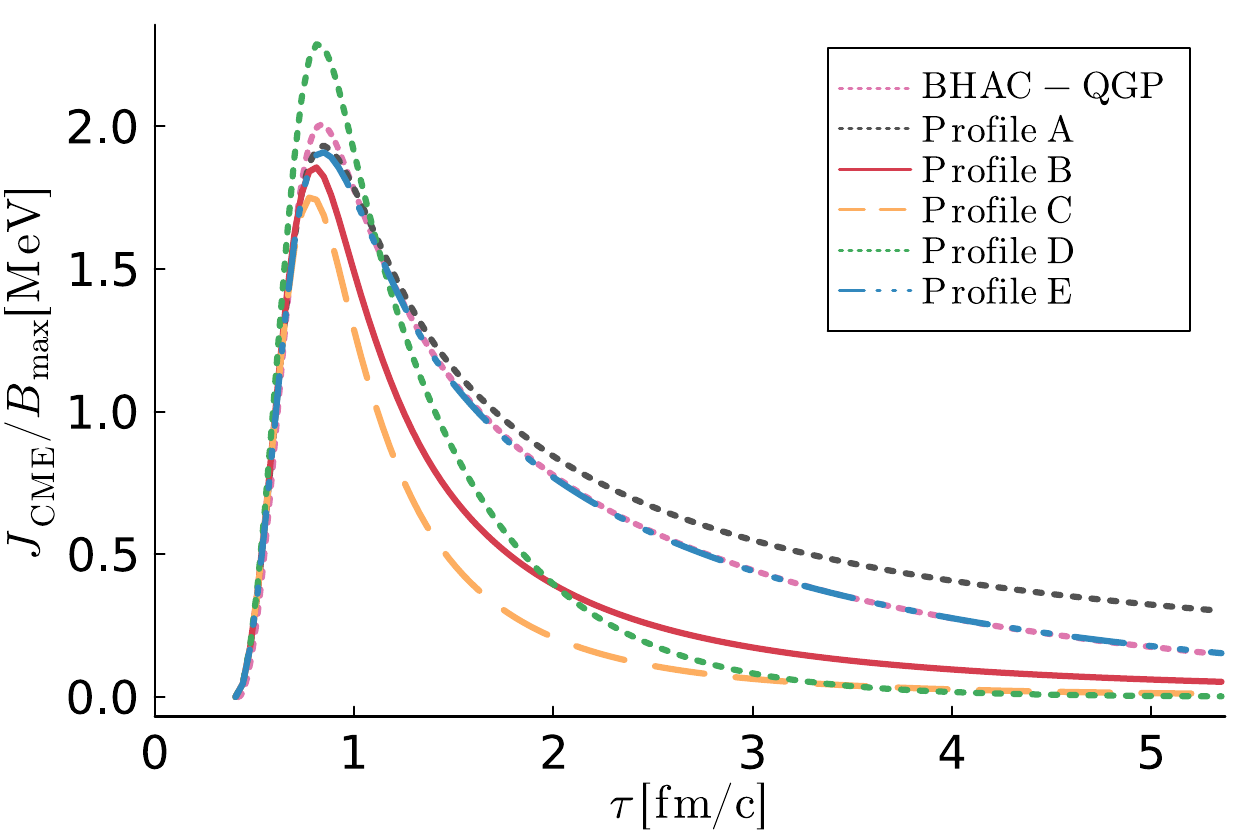} \\ \includegraphics[width=0.85\linewidth]{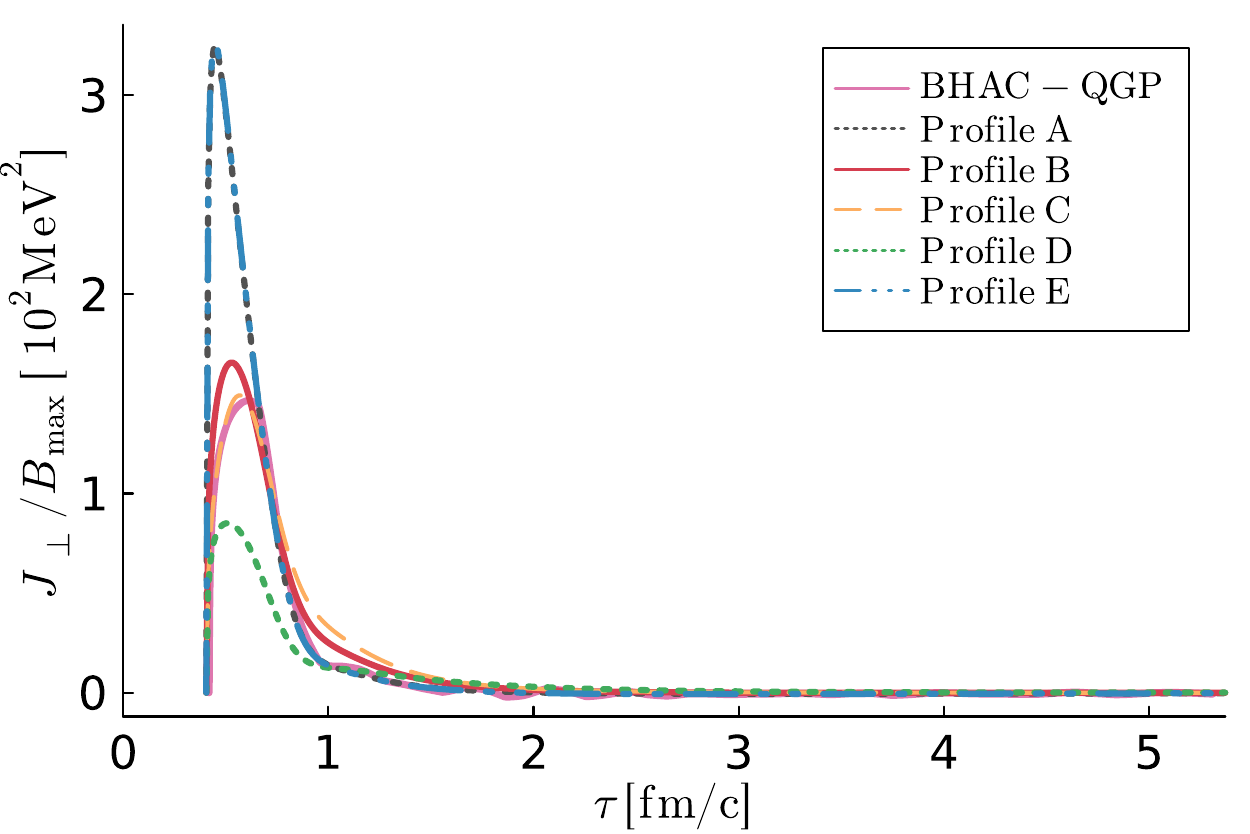}
    \end{tabular}
    \caption{ Normalized chiral magnetic current (top panel) and transverse current (bottom panel) at $\sqrt{s}=200$ GeV for a non-expanding plasma and different magnetic fields (Fig. \ref{fig:magneticfields}). The simulation is initialized at $t_0=0.41$ fm/c.}
   \label{fig:TimeDepenence}
\end{figure}

The CME signal for the time-dependent magnetic fields introduced in Sec. \ref{sec:thermo} is presented in Fig.~\ref{fig:TimeDepenence} (upper panel) for the non-expanding plasma and in Fig. \ref{fig:cme200} (upper panel) for the expanding plasma. At initial time, the plasma has some finite magnetic field and vanishing chiral magnetic current. The response of the system is to build up the CME, a phenomenon that should be independent of the details of the time evolution of the magnetic field provided that the latter is not violently changing. The previous statement is in agreement with the response of the CME in the two figures.

\begin{figure}
    \centering
    \includegraphics[width=0.85\linewidth]{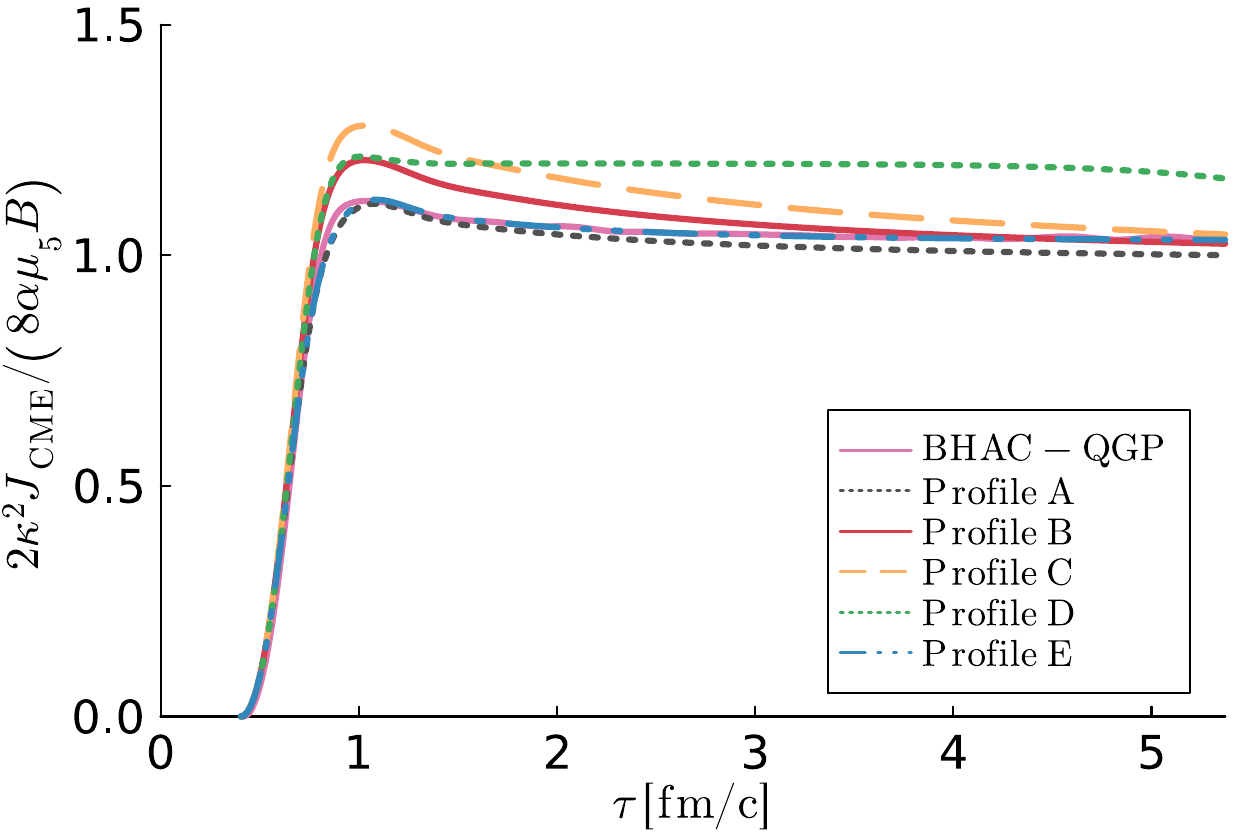}
    \caption{Chiral magnetic current for a non-expanding plasma as in Fig. \ref{fig:TimeDepenence} normalized to the quasi-equilibrium at $\sqrt{s}=200$ GeV. }
    \label{fig:cme_non-exp_quasistatic}
\end{figure}

Once the chiral magnetic current has reached its peak value, it starts to decay. In the non-expanding plasma, the CME signal follows the trend dictated by the time evolution of the magnetic field for profiles A-C, as demonstrated in Fig. \ref{fig:cme_non-exp_quasistatic}. The magnetic field of profiles D and E decay faster than that of profiles A-C, which explains why the chiral magnetic current does not reach the quasi-equilibrium regime. In the expanding plasma, the late-time response of the system also depends on the details of the magnetic field, as discussed in Sec. \ref{sec:inilate}. If the decay of the magnetic field is sufficiently slow, then we reach the quasi-equilibrium regime in which the response of the CME signal is instantaneous and given by the formula \eqref{latecme}, derived for late times and small magnetic fields. This is true for the profiles A, B and C that we study. The agreement between Eq. \eqref{latecme} and the numerical solution is verified in Fig. \ref{fig:cme200late} for such profiles in the expanding case. We can clearly see that the Bjorken expanding magnetic field (profile A) quickly reaches the asymptotic regime, while subleading corrections to the formula \eqref{latecme} delay the onset of quasi-equilibrium for profiles B and C. The magnetic field profiles D, E, $\overline{\rm E}$ and BHAC-QGP decay faster, in the sense specified in Sec. \ref{sec:inilate}, and the quasi-equilibrium formula for the late-time response of the chiral magnetic effect does not apply, also shown in Fig. \ref{fig:cme200late}.

The transverse (circular) currents $J_x$ and $J_y$ in Eq. \eqref{eq:trans_nonexp} are triggered by the time-dependence of the magnetic field itself, and its overall magnitude is controlled by $J_\perp$, defined in Eq. \eqref{eq:jprp}. We show the dynamical evolution of $J_\perp$ in Fig.~\ref{fig:TimeDepenence} (lower panel) for the non-expanding plasma and in Fig.~\ref{fig:cme200} (lower panel) for the expanding counterpart. In the non-expanding case, the initial stages of the evolution are mostly insensitive to the specifics of the dynamics of the magnetic field. After a peak value is reached, the transverse current slowly decreases. Remarkably, the early time-response for the transverse current in the expanding case depends on the chosen profile for the magnetic field, increasing initially for profiles B and C, that decay slower than profile A, while it decreases initially for profiles D, E $\overline{\rm E}$ and BHAC-QGP, which decay faster than profile A. The latter are precisely the profiles for which the late-time formula for the chiral magnetic effect \eqref{latecme} does not apply. The response of the transverse current for profile A is trivially zero. Note that the current generated from the time evolution of the magnetic field is circular and does not lead to accumulation of charge.

\begin{figure}
    \centering
    \begin{tabular}{c}
    \includegraphics[width=0.85\linewidth]{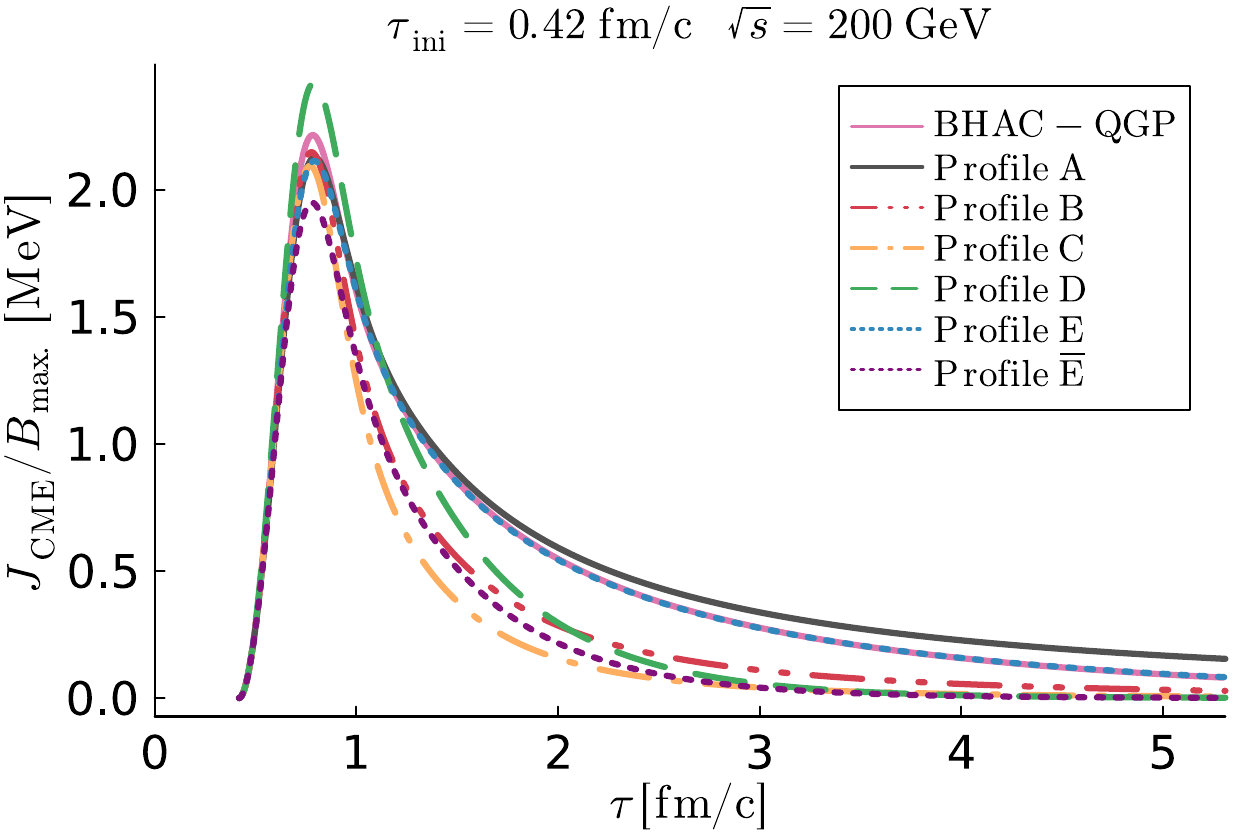} \\ \includegraphics[width=0.85\linewidth]{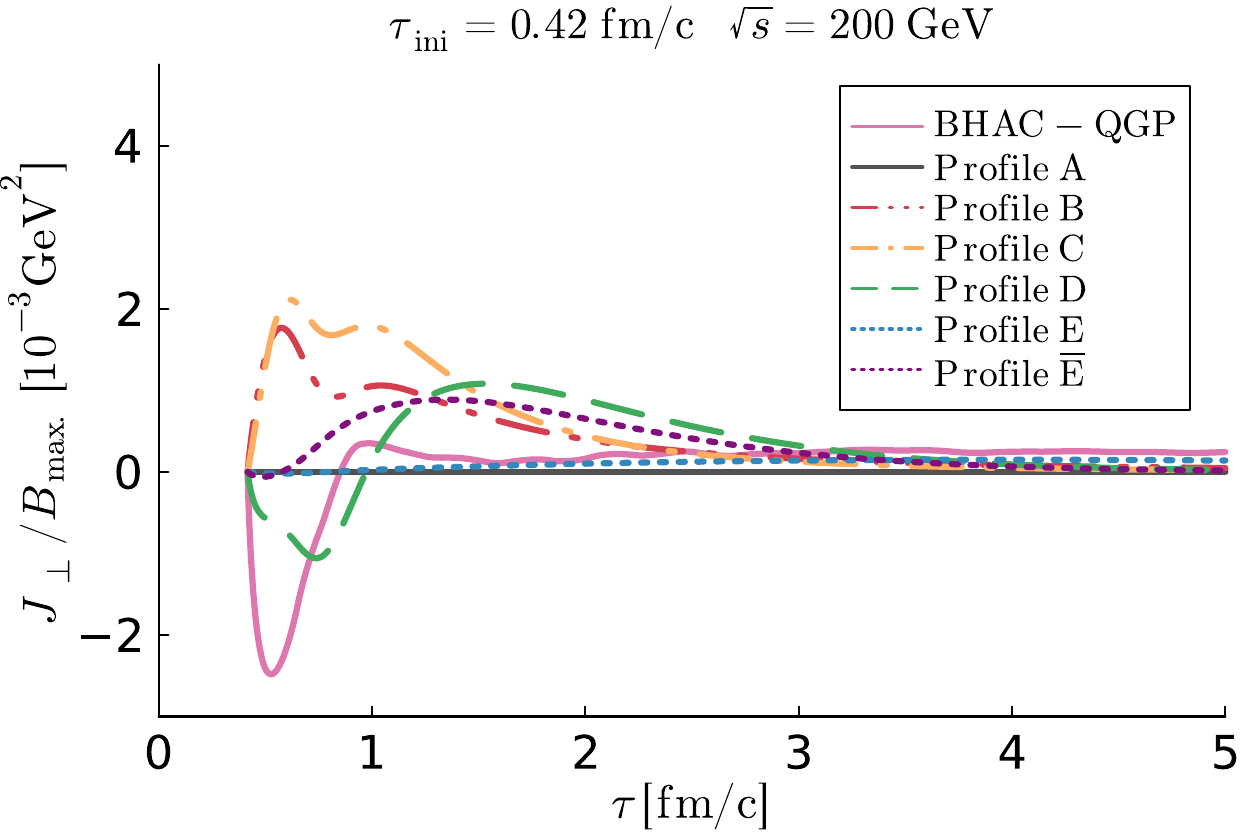}
    \end{tabular}
    \caption{Normalized chiral magnetic current (top panel) and normalized transverse current (bottom panel) at $\sqrt{s}=200$ GeV for an expanding plasma and different time-evolving magnetic fields (see Fig. \ref{fig:magneticfields}). The simulation is initialized at $\tau_{\rm ini} = 0.42$ fm/c.}
    \label{fig:cme200}
\end{figure}
\begin{figure}
    \centering
    \begin{tabular}{c}
    \includegraphics[width=0.85\linewidth]{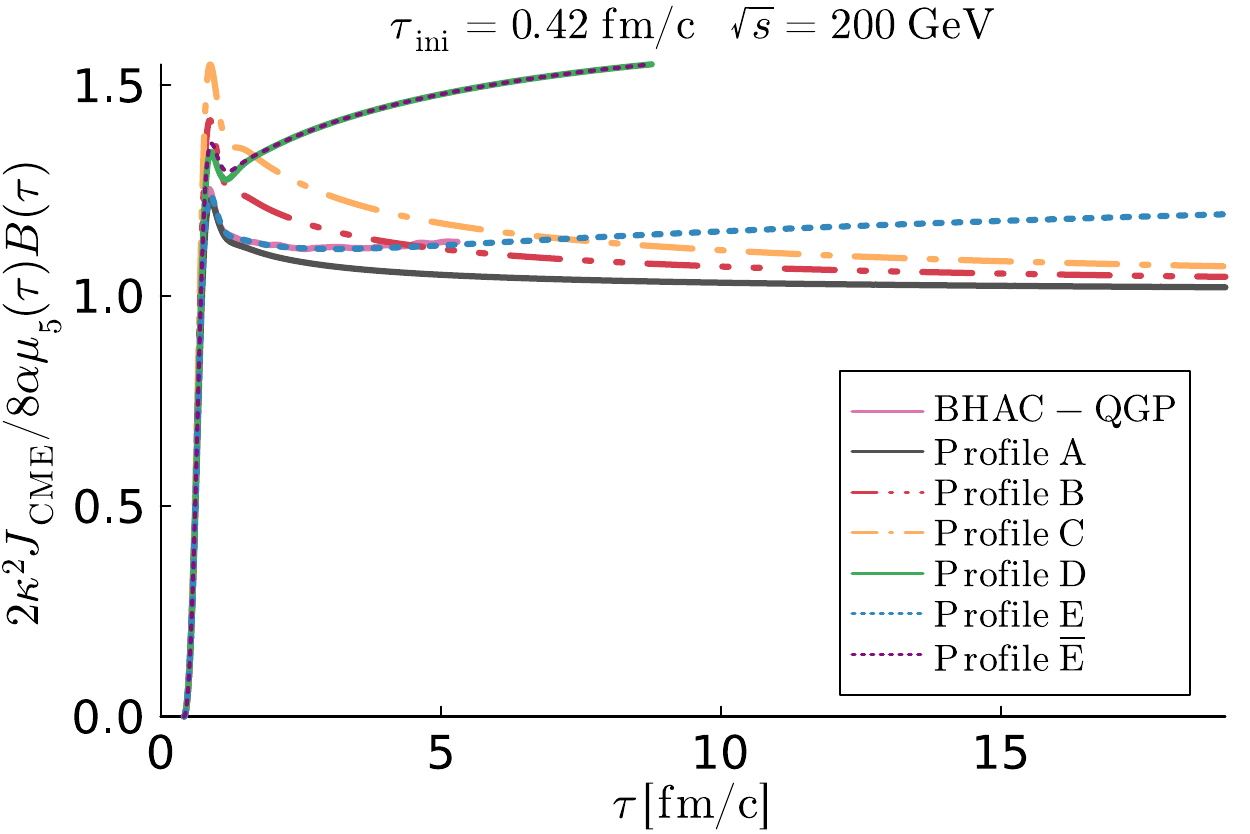}
    \end{tabular}
    \caption{Chiral magnetic current as in Fig. \ref{fig:cme200} normalized to the quasi-equilibrium result of Eq. \eqref{latecme}. The formula applies for the profiles A, B and C, while it does not apply for the other profiles. This is in agreement with the derivation in Sec. \ref{sec:inilate}.}
    \label{fig:cme200late}
\end{figure}

\subsection{Exploring collision energies}\label{sec:energies}

We now turn our attention to the impact of the collision energy $\sqrt{s}$ on the expected signal for the chiral magnetic current. Following Ref. \cite{Cartwright:2021maz}, we take the time-integrated chiral magnetic current normalized to the transverse area as a proxy for the experimentally accessible measurement of the CME. Therefore, we introduce 
\begin{equation}
    q_V\equiv \int dx dy dt \,J_{\rm CME}\,,
\end{equation}
and normalize it to the transverse areas $A = \int dxdy$ and $\tilde{A}=\int d\xi dy$ for the non-expanding (non-exp) and expanding (exp) cases respectively:
\begin{align}\label{eq:integrated}
\dfrac{q_V^{\rm non-exp.}}{A} &= \int_{t_{\rm ini.}}^{t_{\rm end.}} dt \, J^{\rm non-exp.}_{\rm CME.}\,,\nonumber\\ \dfrac{q_V^{\rm exp.}}{\tilde{A}} &=\int_{\tau_{\rm ini.}}^{\tau_{\rm end.}} d\tau \tau \ J^{\rm exp.}_{\rm CME}\,, 
\end{align}
where we have included the superscripts to emphasize that the chiral magnetic current obtained in either case follows from the dynamical evolution of the gauge fields in different background metrics. The initial time for the static scenario is set to $t_{\rm ini.} = 0$ for the profiles B, C and D, while it is set to $t_{\rm ini.} = 0.4$ fm/c in the other cases. Conversely, in the expanding scenario we consider two different initial times, $\tau_{\rm ini}=0.1$ fm/c and $\tau_{\rm ini.}=0.6$ fm/c, as discussed in Sec. \ref{sec:thermo}. In all instances, we set the final time to $t_{\rm end. }=\tau_{\rm end.} = 5$ fm/c.

\begin{figure}
    \centering
    \begin{tabular}{c}
    \includegraphics[width=0.85\linewidth]{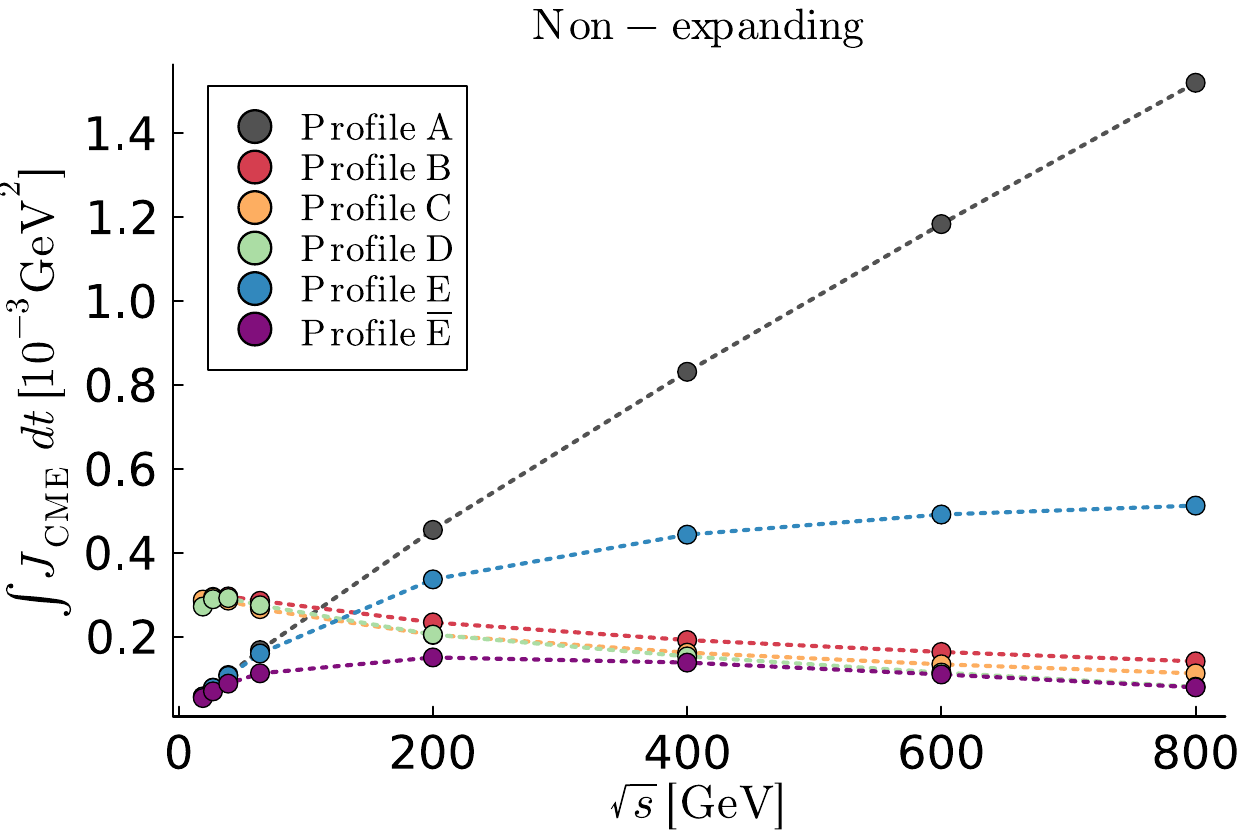} 
    \end{tabular}
    \caption{Integrated measure of the chiral magnetic effect (normalized by unit area) for a non-expanding background as a function of the collision energy. The non-expanding background is introduced in Sec. \ref{sec:non_exp} and the parameters are set according to Sec. \ref{sec:thermo}. We note that the profile A monotonically increases with collision energy. The result for profiles B-D are qualitatively similar and predict an optimal collision energy around $\sqrt{s}=20$ GeV. Profiles E and $\overline{\rm E}$ similarly predict the existence of an optimal collision energy.} 
    \label{fig:integrated-lowenergies}
\end{figure}

\begin{figure}
    \centering
    \begin{tabular}{c}
    \includegraphics[width=0.85\linewidth]{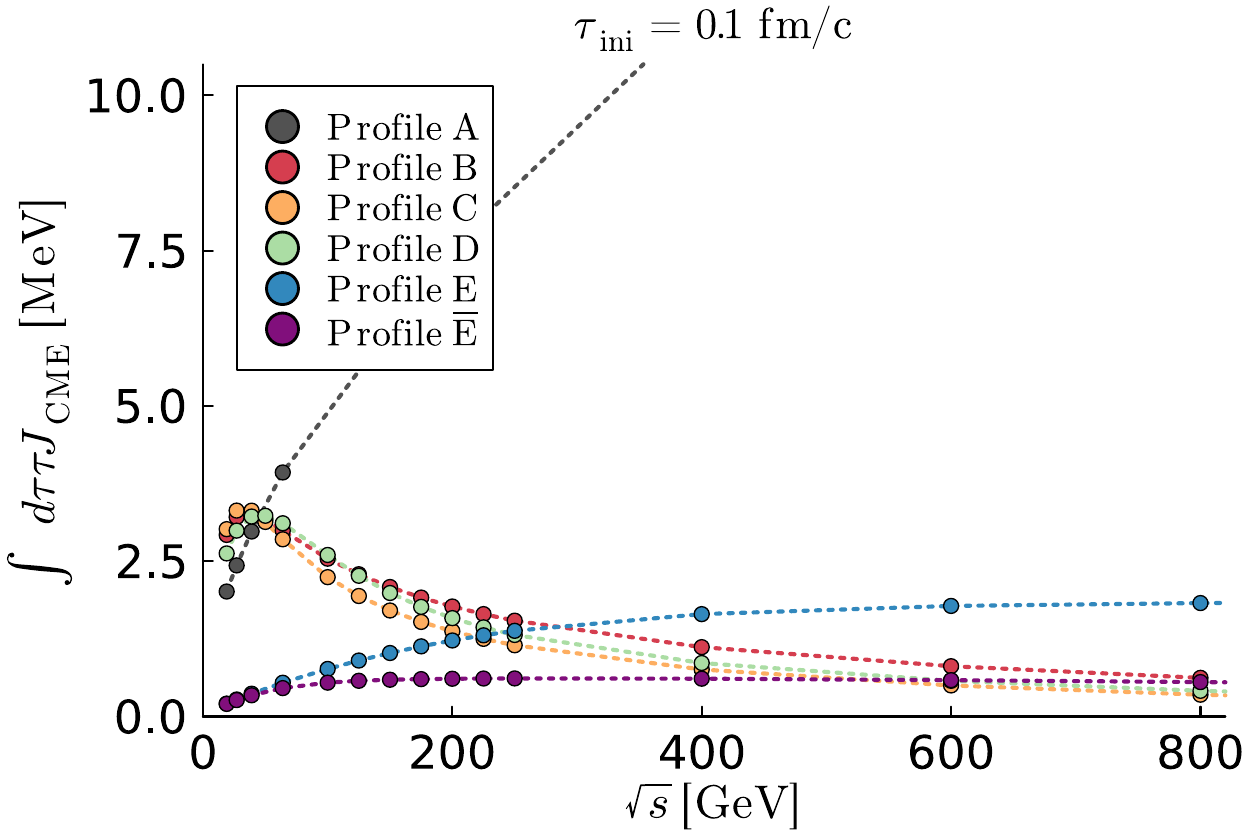} \\ \includegraphics[width=0.85\linewidth]{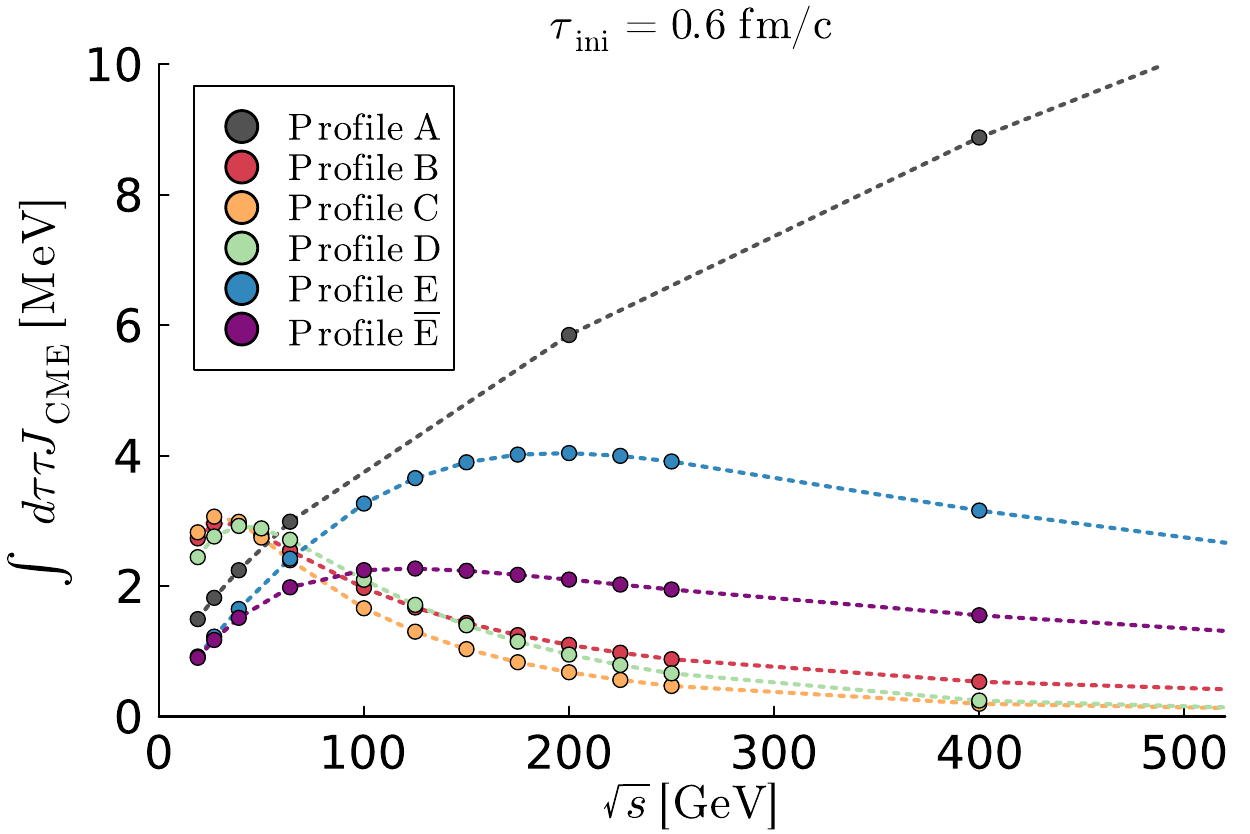}
    \end{tabular}
    \caption{Integrated measure of the chiral magnetic effect (normalized by unit area) for a expanding background as a function of the collision energy for two different choices of initial time $\tau_{\rm ini.}$. The expanding background is introduced in Sec. \ref{sec:expanding_plasma} and the parameters are set according to Sec. \ref{sec:thermo}. We note again that the profile A monotonically increases with collision energy. The result for the profiles B-$\overline{\rm E}$ predict an optimal collision energy below $\sqrt{s}=200$ GeV for which the CME signal is enhanced.} 
    \label{fig:integrated-expanding-lowenergies}
\end{figure}

The results for the integrated measure of the chiral magnetic effect are shown in Fig.~\ref{fig:integrated-lowenergies} for the non-expanding plasma and in Fig. \ref{fig:integrated-expanding-lowenergies} for the expanding plasma at the two quoted values of the initial time $\tau_{\rm ini.}$. The profile A corresponds to the Bjorken expanding magnetic field considered in Ref. \cite{Cartwright:2021maz}. The setup considered in our work corresponds to the Case VI in their work. The authors of \cite{Cartwright:2021maz} conclude that, for Case VI, the chiral magnetic effect signal increases with collision energy, and the same result can be extracted for the Profile A in the figures \ref{fig:integrated-lowenergies} and \ref{fig:integrated-expanding-lowenergies} of our work.

Interestingly, the integrated measure of the chiral magnetic current in the non-expanding plasma for the $1/\sinh(t/t_B)$ extrapolation of the BHAC-QGP data (profiles E and $\overline{\rm E}$), feature a plateau as the energy of the collision is increased. The same holds true for the expanding case with initial time $\tau_{\rm ini}=0.1$ fm/c. Conversely, the profiles B-D in both the non-expanding and the expanding cases predict an optimal collision energy, for which the CME signal is maximal, at approximately $\sqrt{s} \sim 40$ GeV. On the other hand, profiles E and $\overline{\rm E}$ in the expanding scenario with initial time $\tau_{\rm ini.}=0.6$ fm/c suggest that the optimal collision energy lies slightly below $200$ GeV. Finally, the integrated CME signal coming from the Bjorken expanding magnetic field (profile A) is increasing with collision energy without bounds. We emphasize that the BHAC-QGP data of Ref. \cite{Mayer:2024kkv} is given at $\sqrt{s}=200$ GeV, and extrapolating profiles different that the inverse sinh function can lead to different qualitative results. Conversely, the magnetic field profiles B, C and D extracted from \cite{Guo:2019joy} are obtained from a fit to different collision energies $\sqrt{s}\sim10-200$ GeV, and the results displayed in Figs. \ref{fig:integrated-lowenergies} and \ref{fig:integrated-expanding-lowenergies} for those collision energies is more reliable. In all three cases, the integrated measure of the chiral magnetic effect gets enhanced at lower collision energies ($\sqrt{s}<200$ GeV) with the optimal energy around $\sqrt{s} \sim 40$ GeV. The previous conclusion holds true for the profiles B, C and D in the three scenarios considered in this work.

\begin{figure}
    \centering
    \begin{tabular}{c}
    \includegraphics[width=0.85\linewidth]{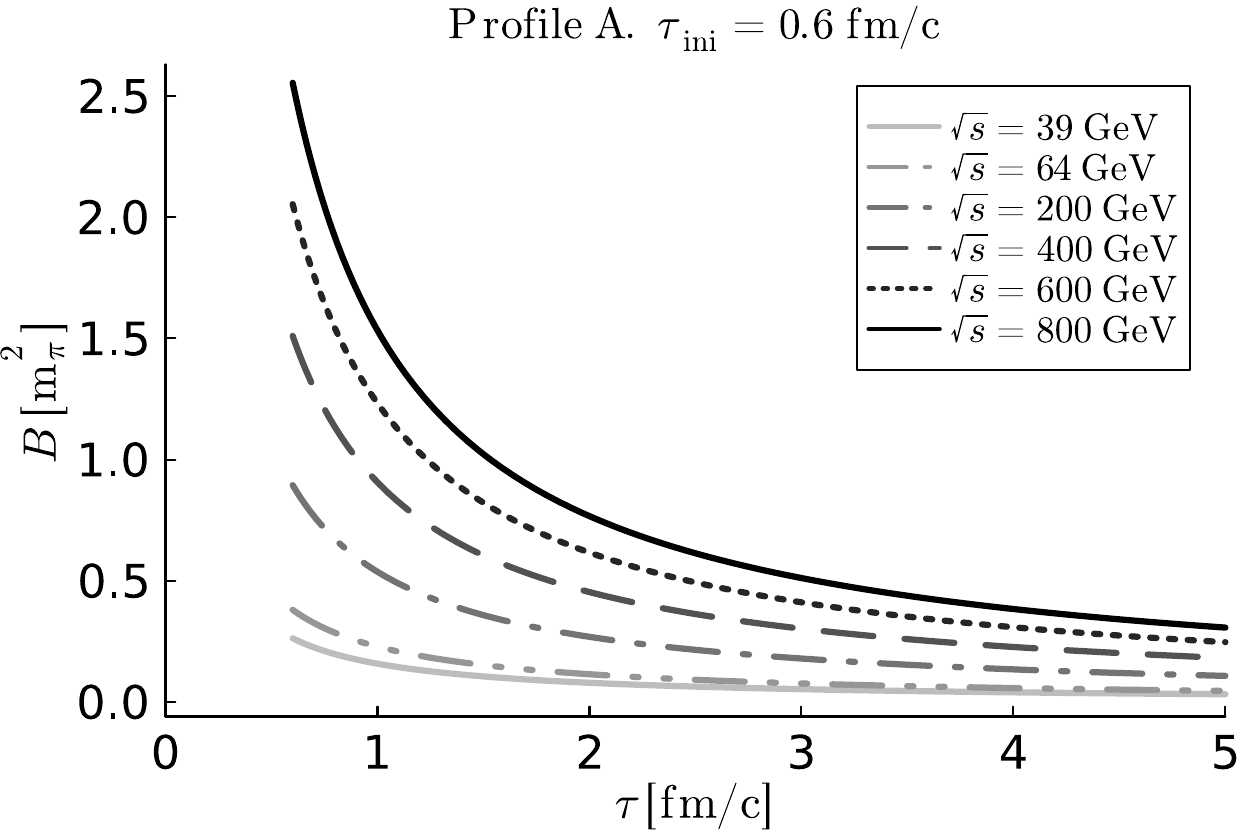} \\ \includegraphics[width=0.85\linewidth]{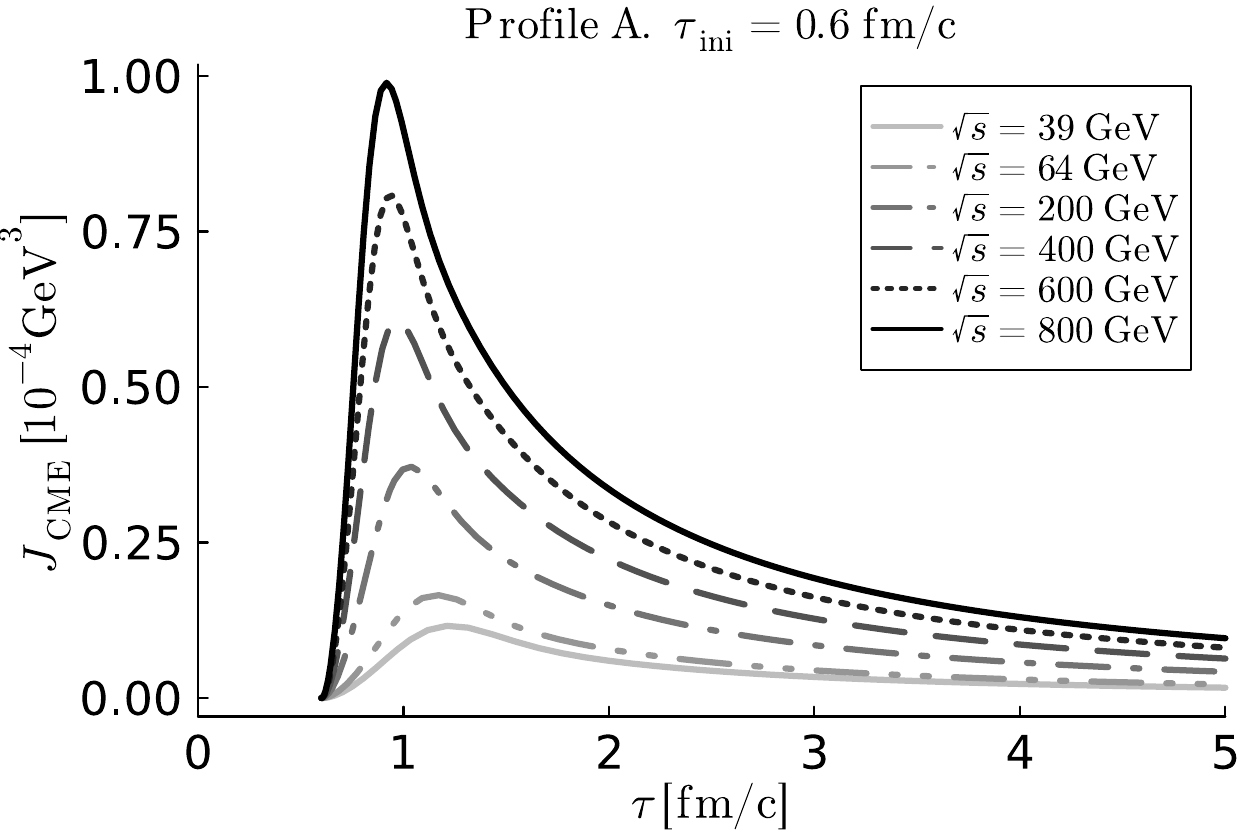}
    \end{tabular}
    \caption{(Top panel) Bjorken expanding magnetic field (profile A in Sec. \ref{sec:thermo}) at different collision energies. (Bottom panel) chiral magnetic current corresponding to the expanding background with profile A for the magnetic field. The CME signal is increased with increasing collision energies.}
    \label{fig:bjork}
\end{figure}

\begin{figure}
    \centering
    \begin{tabular}{c}
    \includegraphics[width=0.85\linewidth]{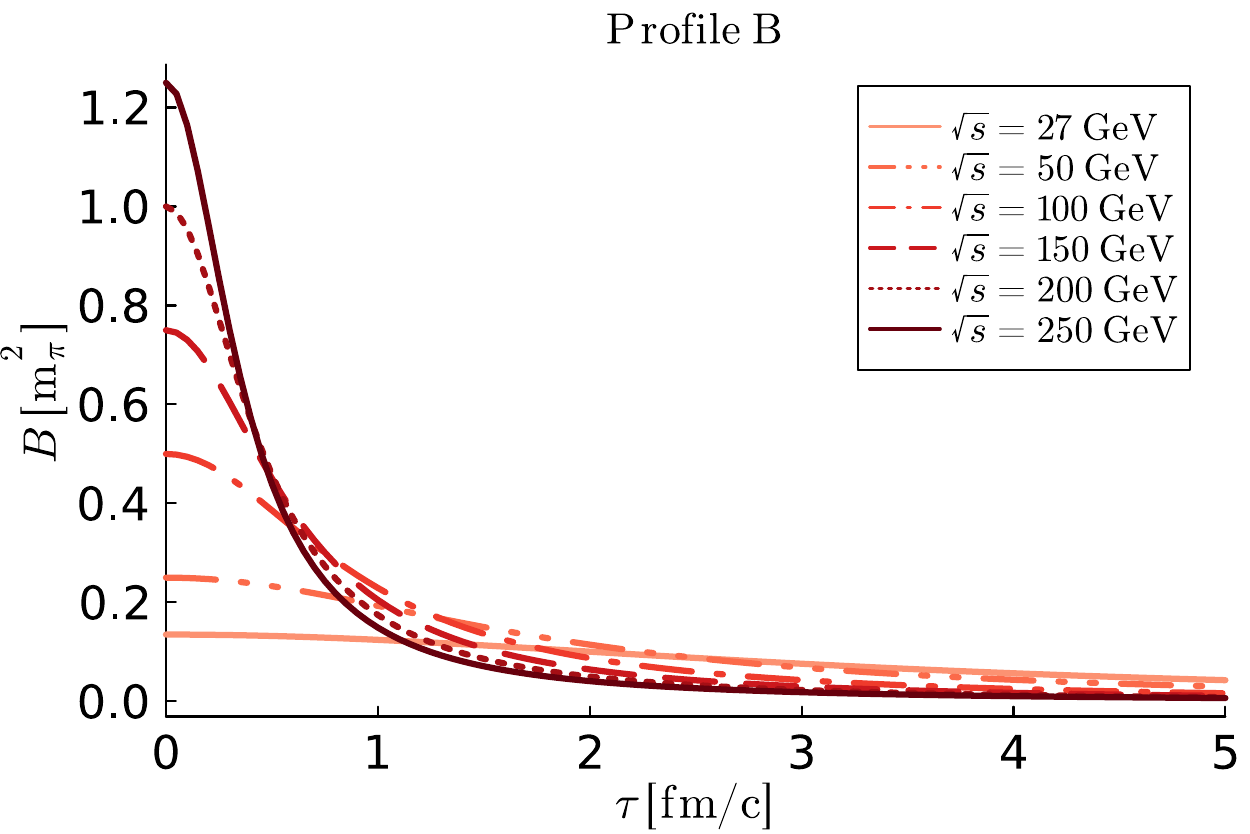} \\ \includegraphics[width=0.85\linewidth]{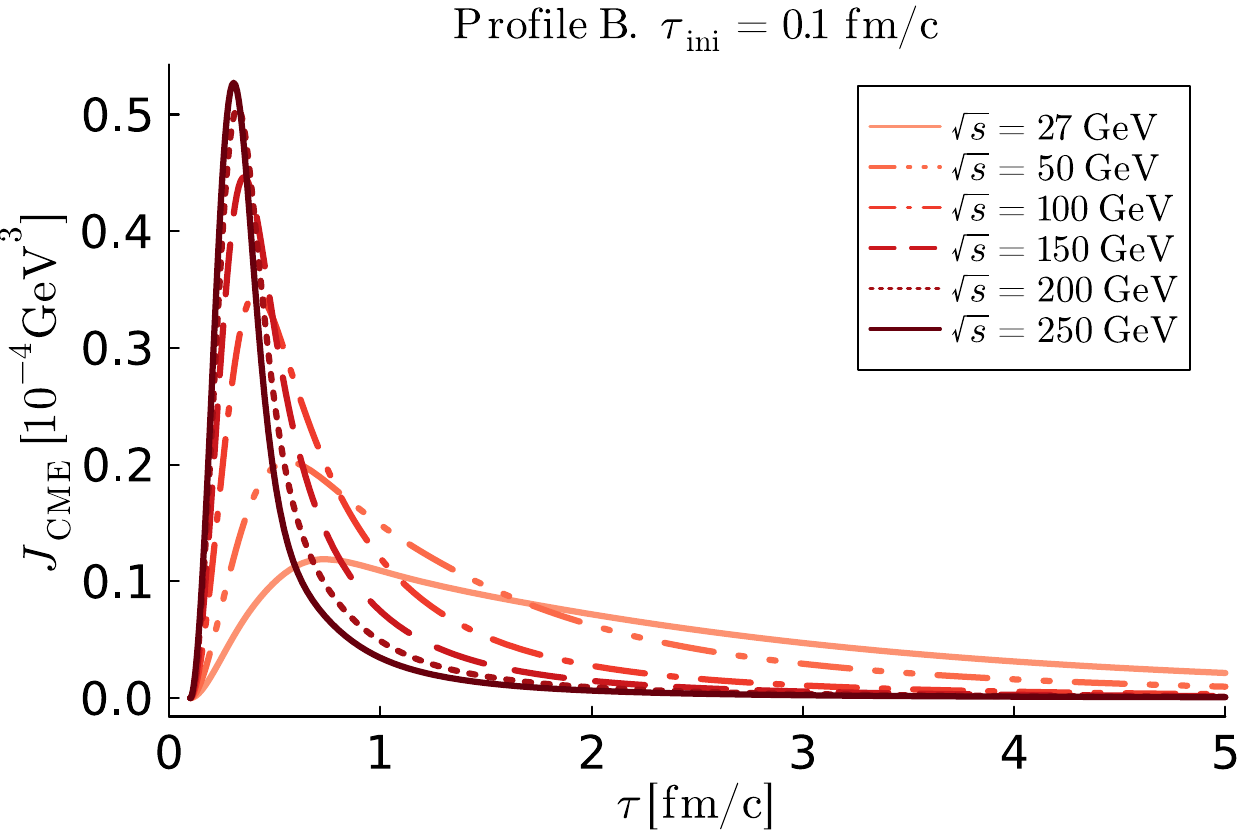} \\ \includegraphics[width=0.85\linewidth]{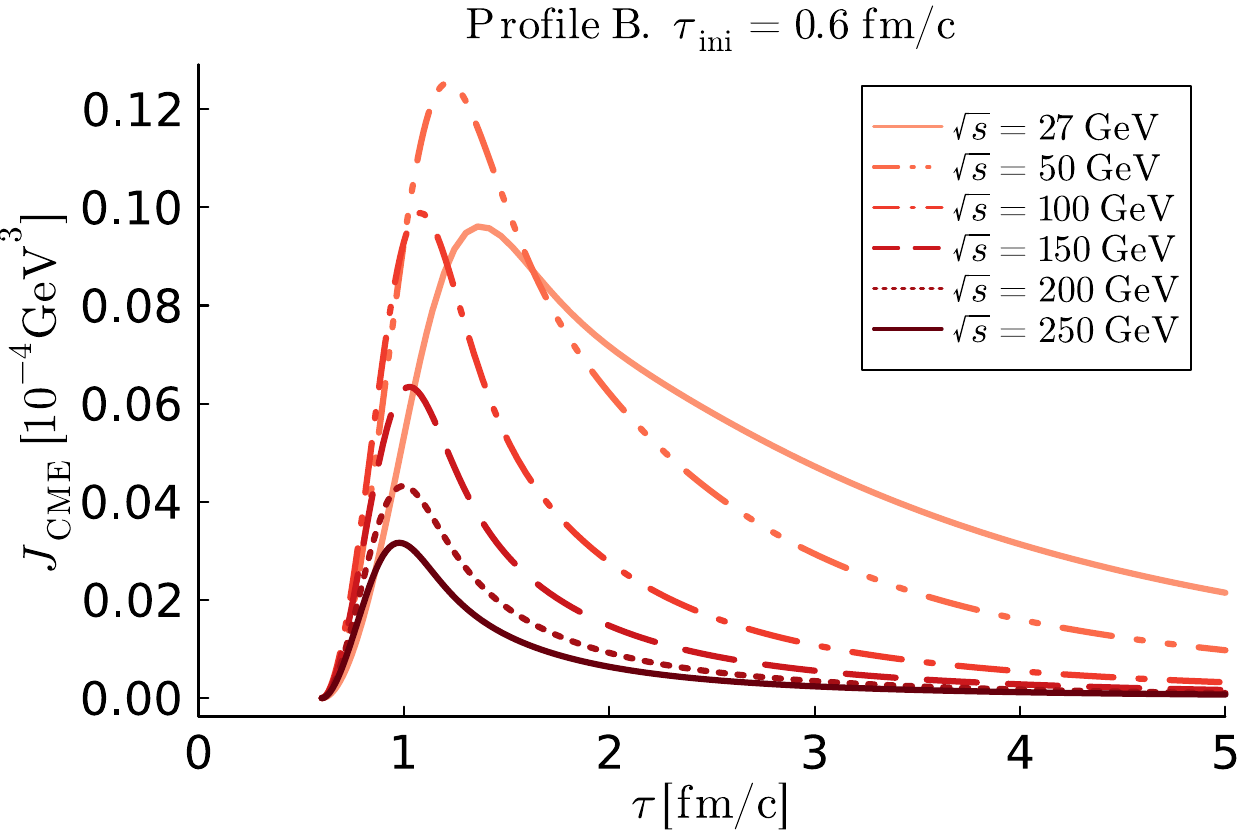}
    \end{tabular}
    \caption{(Top panel) Time dependence of the magnetic field for profile B for different collision energies. (Middle and bottom panels) chiral magnetic current corresponding to the expanding background with the magnetic field displayed in the top panel for two values of the initial time $\tau_{\rm ini.}$. The signal is enhanced at intermediate collision energies and decreases for sufficiently large values of $\sqrt{s}$. }
    \label{fig:canon}
\end{figure}

We now address the discrepancy in the conclusions following from profile A and the rest of the profiles. We argue that the decreasing lifetime of the magnetic field with the energy of the collision, parametrized as $\tau_B\sim(\sqrt{s})^{-1}$, plays a fundamental role. The Bjorken expanding magnetic field (profile A) lacks a measure of the lifetime of the magnetic field, and therefore the scaling of profile A with energy is a trivial increase in the overall magnitude according to Eq. \eqref{B_energy}, as shown in Fig.~\ref{fig:bjork} (top panel). For this reason, it is natural that the chiral magnetic current (Fig.~\ref{fig:bjork} bottom panel) similarly increases in magnitude with the collision energy. Conversely, the profiles B-E of the magnetic field include an extra parameter, $\tau_B$, controlling the different decays of the magnetic field in the plasma. In Fig.~\ref{fig:canon}, we show the results for profile B, as an illustrative example of this discussion, at different collision energies. From the time evolution of the magnetic field (Fig.~\ref{fig:canon} top panel) we already note that, as the energy increases, the magnetic fields cross each other, and the hierarchy of the magnetic fields is inverted. Clearly, this inversion of hierarchies is due to the competing scaling of the magnetic field with energy: the overall peak increases with energy while the lifetime decreases. As a result, the chiral magnetic current, shown in the middle and lower panels of Fig.~\ref{fig:canon} for different initial times $\tau_{\rm ini.}$, also decreases as the energy of the collision is increased, especially at late times. The overall integrated measure, defined in Eq. \eqref{eq:integrated} and shown in Fig.~\ref{fig:integrated-expanding-lowenergies}, is similarly decreasing as the energy of the collision increases. 

\begin{figure}
    \centering
    \includegraphics[width=0.85\linewidth]{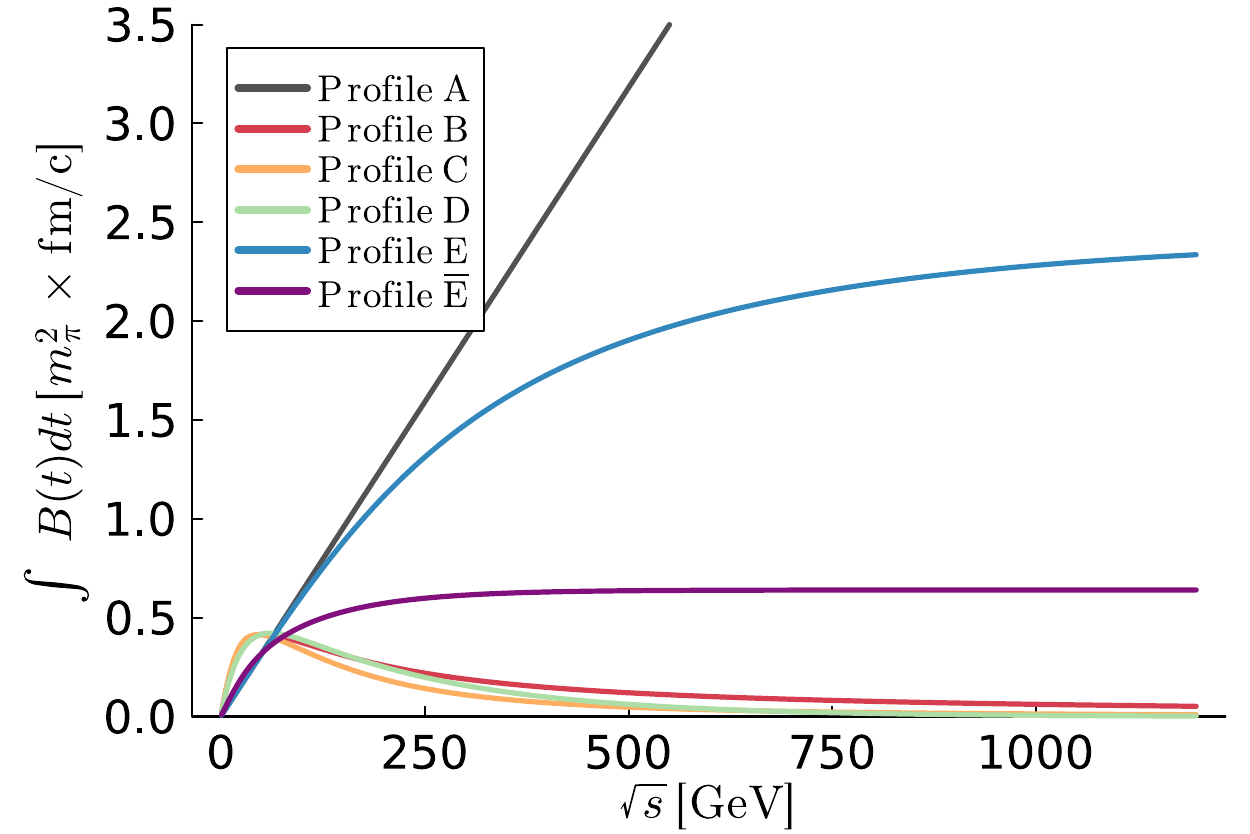}
    \caption{Time integrated magnetic fields between $\tau_{\rm ini} = 0.6$ fm/c and $\tau_{\rm end}=5$ fm/c as a function of the collision energy $\sqrt{s}$. The structure is in close resemblance with the integrated measure of the CME signal in the non-expanding plasma (Fig. \ref{fig:integrated-lowenergies}).}
    \label{fig:integrated_magnetic}
\end{figure}

We note that the energy dependence of the integrated chiral magnetic current is in close resemblance with the integrated magnetic field as a function of the collision energy, see Fig. \ref{fig:integrated_magnetic}. Comparing with the non-expanding results of Fig. \ref{fig:integrated-lowenergies} we recover the existence of an optimal collision energy for profiles B, C, and D; the plateaus for profiles E and $\overline{\rm E}$; and the monotonic increase for profile A. The same structure is observed for the expanding results of Fig. \ref{fig:integrated-expanding-lowenergies}. The absence of the plateau for the expanding case can be attributed to the additional time decay of the effective axial chemical potential and energy density. This result suggests that the energy dependence of the CME signal is largely controlled by the time-dependence of the magnetic field.

We conclude this section commenting on the validity of the model at different collision energies. As we pointed out in Sec. \ref{sec:model}, the model considered here is insensitive to the phase transitions. Moreover, the model is conformal and does not exhibit a realistic scaling behavior of the entropy density when approaching the phase transition. Therefore, the model constitutes a good approximation at high temperatures, similarly high collision energies, where the non-conformal effects are less pronounced. As the collision energy decreases, so does the temperature, and the non-conformal effects become increasingly important, so that the model itself departs progressively from the physical description of the quark-gluon plasma. Nonetheless, the chiral magnetic effect is expected to be augmented slightly above the critical temperature \cite{Fu:2010rs,Fu:2010pv}, and we expect that the enhancement of the chiral magnetic effect at smaller collision energies to similarly hold for the physical quark-gluon plasma.

\section{Discussion and Conclusions}\label{sec:conclusions}

The uncertainties related to the dynamical evolution of the quark-gluon plasma, generated in heavy-ion collisions, hinder an accurate theoretical description of the expected chiral magnetic effect signal. In this paper, we focus on the impact of the dynamical evolution of the magnetic field on the out-of-equilibrium response of the chiral magnetic effect in the QGP. To this end, we use a holographic model, introduced in \cite{Gynther:2010ed}, which provides a window into the strongly coupled out-of-equilibrium dynamics of the CME in a conformal plasma. 

In Ref.~\cite{Ghosh:2021naw} a similar question was addressed for a constant electromagnetic field in a non-expanding plasma, while in Ref.~\cite{Cartwright:2021maz} the discussion was extended to a purely Bjorken expanding plasma, focusing on the energy dependence. Furthermore, in \cite{Grieninger:2023myf} we extended this study to include the effects of the non-abelian anomaly which are responsible for the occurrence of the chiral magnetic effect in QCD. The three studies considered a full backreacted geometry, without any assumption on the strength of the magnetic field. We partially extend the previous constructions to include a more generic time dependence of the magnetic field, while assuming that the magnetic field is small compared to the energy density of the plasma. The validity of the small magnetic field assumption is discussed in Sec.~\ref{sec:model}~(Fig. \ref{fig:comparison}). This assumption is technically convenient, because the transverse-plane anisotropies due to the time-dependence of the magnetic field can be neglected to leading order. 

The extension to time-dependent magnetic fields is done for both a non-expanding and an expanding plasma. We obtain the ``late-time" geometry of the chirally charged expanding plasma, as done in Ref.~\cite{Kalaydzhyan:2010iv}, and we are able to derive analytically the (quasi)-equilibrium formula for the chiral magnetic effect in the expanding plasma:
\begin{equation}\label{cmelateconc}
    2\kappa^2 J^{\text{ quasi-eq}}_{\textrm{CME}} =8\alpha\mu_5 B_z\,,
\end{equation}
where an effective notion of the axial chemical potential in this quasi-equilibrium setup is defined in analogy to the non-expanding equilibrium state [see Eq. \eqref{mu5exp}]. The same ``late-time" expanding geometry is later used as a background to study the evolution of the CME.

The question of the dynamical evolution of the magnetic field in the quark-gluon plasma is a research topic in its own right, with several research groups devoted to its understanding and description. In order to ameliorate the uncertainty in the CME coming from the dynamical evolution of the magnetic field, and extract more robust conclusions, we incorporate in our description the different profiles $B(\tau)$ studied in Refs.~\cite{Guo:2019joy,Mayer:2024kkv}, see also Fig.~\ref{fig:magneticfields}. On the other hand, we choose the values of temperature and axial charge as in Ref.~\cite{Cartwright:2021maz}, and we similarly study the energy dependence of the integrated CME signal. Note that the work of Ref.~\cite{Cartwright:2021maz} uses a purely Bjorken expanding magnetic field (profile A in our work). Finally, we consider two values of the initial time, $\tau_{\rm ini.} =0.1$ fm/c and   $\tau_{\rm ini.} =0.6$ fm/c. The first of them encodes the fact that the initial stages of the plasma are gluon dominated and there should be no response from the quarks \cite{Huang:2022qdn}, while the second is a more conservative approach that initializes the evolution at the equilibration time. 

In Sec.~\ref{sec:RHIC}, we study the sensitivity of the chiral magnetic current to the different possible time evolution of the magnetic field. It is manifest that the qualitative features of the CME are similar to each other. At early times, the current builds up independently of the details of the magnetic field evolution. Shortly afterwards, there is a transient regime where the CME reaches its maximum value. Finally, the current decays following the profile of the magnetic field in agreement with Eq.~\eqref{cmelateconc}. The same conclusions hold for the expanding and non-expanding scenarios. In addition, we note that the time gradients of the homogeneous magnetic field generate a circular current $J_\perp$, which peaks shortly after the start of the simulation and disappears at late times.  

We then turn our attention to the energy dependence of the integrated CME in Sec.~\ref{sec:energies}. We recover the result obtained in Ref.~\cite{Cartwright:2021maz} (case VI in their work) that, for profile A, increasing the collision energy leads to an increase in the integrated CME observable without bounds. However, the conclusion is the opposite for all other profiles (B-$\overline{\rm E}$) considered in our work, since we observe that the integrated measure of the chiral magnetic current is generally non-monotonic with collision energy. The non-monotonicity of the integrated CME signal is encompassed by the prediction of an optimal collision energy for which the observed signal is maximal. The exact value of the optimal collision energy depends on the studied magnetic field profile. We argue that the source of the discrepancy between the conclusions following from profile A and profiles B-$\overline{\rm E}$ stems from the fact that profile A has no parameter controlling the lifetime of the magnetic field at different collision energies, while the profiles B-$\overline{\rm E}$ take into account that the lifetime of the magnetic field decreases with increasing collision energy. Note that in all cases, the peak value of the magnetic field increases with the energy of the collision. We further conclude that the energy dependence of the integrated chiral magnetic current is dominated by the integrated magnetic field in each case.

The previous discussion highlights the need to include the time evolution of the magnetic field in the holographic simulations describing the out-of-equilibrium evolution of the chiral magnetic current. The different cases studied in this work also lead us to conclude that the interplay between the lifetime of the magnetic field and its peak value leads to an enhancement of the CME signal at lower collision energies. The magnetic field profiles B, C and D are the only ones obtained from a fit at different collision energies, and therefore the extrapolation to different collision energies is more reliable in these three cases. The profiles B, C and D predict that the optimal collision energy is about $\sqrt{s}\sim 40$ GeV, with the integrated CME signal decreasing as the collision energy exceeds this value.

We also find that the inclusion of the non-conformal effects can become relevant as we approach the phase transition temperature from above. While we expect the conclusions of this work to remain qualitatively unchanged, a separate study is required to assess the effects of the non-conformality of the plasma. This study could be conducted using a V-QCD model similar to the one used in~\cite{Gallegos:2024qxo}. Other avenues for future research include the study of the inhomogeneous effects that appear from the dependence of the magnetic field on the transverse plane coordinates, which can be readily studied in the model used in this paper. In the discussion so far, we have considered the magnetic field as a source, whose dependence on time and transverse coordinates is fixed in advance. The incorporation of the magnetic fields in a truly dynamical approach is an interesting topic that should also be studied separately. The work of Ref. \cite{Baggioli:2024zfq} already provides the setup for such a dynamical study. 

Finally, from a theoretical point of view, it would be very interesting to better understand the black hole geometries subjected to a very strong magnetic field in the presence of anomalies. In the work of~\cite{Grieninger:2016xue,Ammon:2016fru} (and subsequently \cite{Haack:2018ztx, Ghosh:2021naw,Grieninger:2023myf, Grieninger:2023wuq}) observed undamped oscillations of the chiral magnetic current that can be traced back to a quasi-normal mode approaching the real axis (for increasing magnetic field) in the complex frequency plane. This was also recently discussed in \cite{Meiring:2023wwi,Waeber:2024ilt}. These undamped oscillations only appear when the Chern-Simons coupling (strength of the abelian anomaly) is above a certain critical value, which seems to be in agreement with Ref.~\cite{DHoker:2009ixq}. It would also be interesting to investigate how time-dependent magnetic fields affect this behavior.

\section*{Acknowledgments}

We thank A. Bandyopadhyay for useful discussions. SG was supported in part by a Feodor Lynen Research fellowship of
the Alexander von Humboldt foundation. SMT was supported by the EU’s NextGenerationEU instrument through the National Recovery and Resilience Plan of Romania - Pillar III-C9-I8, managed by the Ministry of Research, Innovation and Digitization, within the project entitled ``Facets of Rotating Quark-Gluon Plasma'' (FORQ), contract no. 760079/23.05.2023 code CF 103/15.11.2022. The work of P.G.R. is supported through the grants CEX2020-001007-S and PID2021-
123017NB-100, PID2021-127726NB-I00 funded by MCIN/AEI/10.13039/501100011033 and
by ERDF “A way of making Europe”

\bibliography{main}

\end{document}